%% file: Poiseuille_curved_space_v6.3.tex
\documentclass[aps, pre, twocolumn,showkeys,showpacs,amsmath,amssymb,superscriptaddress]{revtex4-1}

\input{pakete}
\input{befehle}
\input{makros}

\begin{document}

\title{Poiseuille flow in curved spaces}

\author{J.-D. Debus} \email{debusj@ethz.ch} \affiliation{ ETH
  Z\"urich, Computational Physics for Engineering Materials, Institute
  for Building Materials, Wolfgang-Pauli-Strasse 27, HIT, CH-8093 Z\"urich,
  Switzerland}
  
\author{M. Mendoza} \email{mmendoza@ethz.ch} \affiliation{ ETH
  Z\"urich, Computational Physics for Engineering Materials, Institute
  for Building Materials, Wolfgang-Pauli-Strasse 27, HIT, CH-8093  Z\"urich,
  Switzerland}

\author{S. Succi} \email{succi@iac.cnr.it} \affiliation{Instituto per le Applicazioni del Calcolo C.N.R., Via dei Taurini, 19 00185, Rome, Italy}
  
\author{H. J. Herrmann}\email{hjherrmann@ethz.ch} \affiliation{ ETH
  Z\"urich, Computational Physics for Engineering Materials, Institute
  for Building Materials, Wolfgang-Pauli-Strasse 27, HIT, CH-8093 Z\"urich,
  Switzerland}

\begin{abstract}

We investigate Poiseuille channel flow through intrinsically curved  media, equipped with localized metric perturbations. To this end, we study the flux of a fluid driven through the curved channel in dependence of the spatial deformation, characterized by the parameters of the metric perturbations (amplitude, range and density). We find that the flux depends only on a specific combination of parameters, which we identify as the average metric perturbation, and derive a universal flux law for the Poiseuille flow.
For the purpose of this study, we have improved and validated our recently developed lattice Boltzmann model in curved space by considerably reducing discrete lattice effects. 

\pacs{47.11.-j, 02.40.-k}
\end{abstract}

\maketitle

\section{Introduction}\label{sec:introduction}

In many physical systems in Nature, the motion of particles is restricted to a given manifold (e.g., to a two-dimensional surface). For example, soap bubbles are commonly described as a two-dimensional fluid moving on a spherical surface, since the perpendicular radial dimension is negligibly small \cite{seychelles2008}. 
The same concept applies to curved two-dimensional interfaces, e.g., molecular films around an emulsion or aerosol droplet, soap films or foam bubbles, the Earth's atmosphere or the photosphere of the sun \cite{scriven1959dynamics, danov2000viscous, chomaz2001dynamics,priest1982,cao1999navier}.
Because of their experimental accessibility, soap films are of particular interest for the study of 2D surface flow \cite{reuther2015interplay}, as they can be fabricated on large scales \cite{zhang2000flexible} and provide the possibility to study custom-made curvature effects on the flow experimentally. 
Another important field of application is the dynamics of lipid bilayers in microbiology \cite{arroyo2009relaxation}. These lipid bilayers are of particular importance since they constitute the envelope of most of the cell components. In particular, lipid bilayer membranes are known to enter a fluid phase above a transition temperature \cite{arroyo2009relaxation} and can thus be described as a viscous two-dimensional fluid moving along the curved membrane \cite{barrett2015numerical, hu2007continuum, fan2010hydrodynamic}, as has been confirmed experimentally \cite{dimova2006practical, cicuta2007diffusion, amar2007stokes}. Besides that, electron transport on 2D graphene sheets is known to follow the Navier-Stokes equation in the hydrodynamic regime \cite{torre2015nonlocal,mendoza2011preturbulent}. Since graphene sheets can form ripples \cite{fasolino2007intrinsic}, a natural extension to describe electron flow on curved graphene sheets is the Navier-Stokes equation in curved space.
Thinking of cosmological scales, space-time itself can be seen as a four-dimensional, intrinsically curved medium whose curvature is caused by the matter it contains according to Einstein's field equations \cite{landau1975}. As a consequence, flow of cosmic dust is deflected in the gravitational and space-curving field of the surrounding stars.

\begin{figure}[t]
\includegraphics[width=.6\columnwidth]{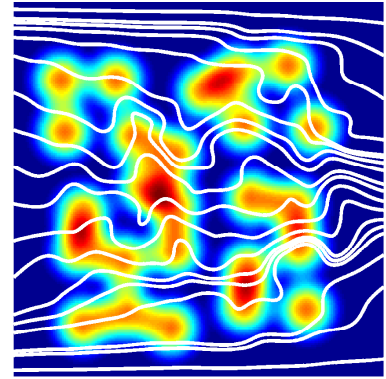}
\caption{Velocity streamlines of a fluid driven through a curved medium with randomly distributed metric perturbations. The colors represent the strength of the perturbation $\dg$ added to the diagonal components of the metric tensor, ranging from flat space (blue) to high perturbation (red).}
\label{fig:paper_title_picture}
\end{figure}

Recently, we have found that local sources of curvature induce dissipation within a viscous fluid \cite{debus2015}. In this paper, we study a different setup, namely Poiseuille-like channel flow through an intrinsically curved space, where the dissipation exerted by the walls dominates over the dissipation induced by the metric perturbations, and therefore, the latter can be neglected. 
Starting from flat space, we introduce local sources of curvature by adding localized perturbations $\dg$ to the Riemann metric $g$. Note that these perturbations do not mimic solid obstacles or boundary walls, since the curved regions of space are still permeable to the flow and thus cannot mimic no-slip boundary conditions. Instead, the metric perturbations lead to an intrinsic curvature of space.
The fluid is driven through the curved channel according to the Navier-Stokes equations in curved space, as illustrated in Fig. \ref{fig:paper_title_picture}.
For different configurations of metric perturbations, we measure the flux of the Poiseuille-like flow in the stationary state and establish a connection between the flux and the curvature parameters. To this end, we consider a two-dimensional channel with regularly and randomly arranged metric perturbations (see Figs. \ref{fig:metric_regular} and \ref{fig:metric_random} for a sketch), which are characterized by their amplitude $a_0$, range $r_0$, and number density $n$. We find that the flux depends only on the specific combination of parameters, which we identify as the average metric perturbation $\langle \dg \rangle \sim a_0 r_0^2 n$. Plotting the flux as function of $\langle \dg \rangle$, we observe that all the data points collapse onto a single curve, which we describe by an empirical flux law.
Secondly, we investigate flow through a three-dimensional curved channel, in order to clarify the influence of the spatial dimension. In three dimensions, we find an analogous flux law to the two-dimensional case.

For the simulations, we use the lattice Boltzmann (LB) method, which describes the motion of a fluid from the perspective of kinetic theory (for a review of the standard LB method see Ref. \cite{succi2015lattice,benzi1992lattice,chen1998}). Recently, the LB method has been extended to general Riemannian manifolds \cite{mendoza2013, mendoza2014, debus2014}, allowing us to simulate fluid flow in nearly arbitrary, smooth geometries. 
Here, we have further improved the method in Ref. \cite{debus2014} by improving the treatment of discrete lattice effects, which originate from the forcing term of the LB equation and which lead to spurious source terms in the Navier-Stokes equations (see also Ref. \cite{Guo2002}). By significantly reducing the discrete lattice effects, we have been able to improve the accuracy of our method considerably.

The paper is organized as follows:
In Sec. \ref{sec:LB-method} we review the lattice Boltzmann method in curved space and present an improved model in which spurious discrete lattice effects are canceled at order $\dt$.
Next, in Sec. \ref{sec:campylotic-media}, we study channel flow through different types of curved media, starting from two-dimensional media with regularly and randomly arranged metric perturbations and ending with three-dimensional media. We compare the results to a previous study in Ref. \cite{mendoza2013}.
Finally, we present a validation of the model by showing the improvement of the method with respect to discrete lattice effects and by confirming that our results are not biased by finite resolution effects.

\section{Lattice Boltzmann Method}\label{sec:LB-method}

\subsection{Review: Lattice Boltzmann in curved space}

In this section, we present a short summary of the method used to simulate Poiseuille flow on Riemann manifolds, equipped with a stationary Riemann metric $g$. More details and validations of the method can be found in previous publications \cite{mendoza2013,mendoza2014,debus2014}. We will use the following notation for partial derivatives throughout the whole paper:
\begin{align*}
	\del_t := \pdiff{}{t},\quad
	\del_i := \pdiff{}{x^i}.
\end{align*}
Furthermore, we use the Einstein sum convention, i.e. repeated indices are summed over.
At first, we specify a $D$-dimensional coordinate system by choosing local coordinates $(x^1,x^2,...,x^D)$ with the corresponding standard vector basis $(\vec e_1, \vec e_2, ...,\vec e_D) = (\frac{\del}{\del x^1}, \frac{\del}{\del x^2}, ..., \frac{\del}{\del x^D})$. In the following, all vectorial and tensorial quantities will be expressed in this basis, e.g. the components of the Riemann metric tensor $g$ are given by $g_{ij} = g(\vec e_i, \vec e_j)$.

The basic equation in our model is the lattice Boltzmann equation in curved space, 
\begin{align}
	\label{eq:LB}
	f_\l(\vec r + \vec c_\l \dt, t + \dt) - f_\l(\vec r, t) 
	 = - \frac{1}{\t} \left( f_\l - f_\l^{\rm eq} \right) + \dt
	 \mathcal F_\l,
\end{align}
which describes the evolution of a distribution function $f_\l(\vec r, t)$ on a lattice. Here $\vec r = (x^1, x^2, ..., x^D)$ denotes the lattice position and $\l$ labels the discrete lattice velocities corresponding to the velocity vectors $\vec c_\l$. We use the Bhatnagar-Gross-Krook (BGK) approximation \cite{BGK} for the description of the particle collisions, meaning that all collisions are characterized by a single-relaxation time $\t$. The Maxwell-Boltzmann equilibrium distribution is denoted by $f_\l^{\rm eq}$, and $\mathcal F_\l$ represents a forcing term, which accounts for the inertial forces in a curved space. Since we perform simulations in both two- and three-dimensions, we choose the $D3Q41$ lattice \cite{chikatamarla2009}. By using this special lattice, the moments of the distribution function are exactly preserved up to third order due to the properties of Gauss-Hermite quadrature (see Appendix \ref{sec:app1}). The lattice contains 41 discrete velocity vectors $\{\vec c_\l\}$ (see Table \ref{tab:weights}) and the speed of sound is given by $c_s^2 = 1-\sqrt{2/5}$. 
\begin{table}
  \centering 
  \begin{tabular}{|@{\quad}c@{\quad}|@{\quad}c@{\quad}|@{\quad}c@{\quad}|}\hline
    $\l$ & $\vec c_\l$ & $w_\l$ \\[2pt] \hline\hline
    1 & $(0,0,0)$ & $\frac{2}{2025} \left(5045-1507\sqrt{10}\right)$ \\[2pt] \hline
    2,3 & $(\pm 1,0,0)$ &  \\ 
    4,5 & $(0,\pm 1,0)$ & $\frac{37}{5\sqrt{10}} - \frac{91}{40}$ \\ 
    6,7 & $(0,0,\pm 1)$ &  \\[2pt] \hline
    8-11 & $(\pm 1,\pm 1, 0)$ &  \\ 
    12-15 & $(\pm 1,0,\pm 1)$ & $\frac{1}{50} 	\left(55-17\sqrt{10}\right)$ \\ 
    16-19 & $(0,\pm 1,\pm 1)$ &  \\[2pt] \hline
    20-27 & $(\pm 1,\pm 1,\pm 1)$ & $\frac{1}{1600}
    \left(233\sqrt{10}-730\right)$ \\[2pt] \hline 
    28,29 & $(\pm 3,0,0)$ &  \\ 
    30,31 & $(0,\pm 3,0)$ & $\frac{1}{16200} \left(295-92\sqrt{10}\right)$ \\ 
    32,33 & $(0,0,\pm 3)$ &  \\[2pt] \hline
    34-41 & $(\pm 3,\pm 3,\pm 3)$ & $\frac{1}{129600}
    \left(130-41\sqrt{10}\right)$ \\[2pt] \hline
  \end{tabular}
  \caption{Discrete velocity vectors $c_\l$ of the D3Q41 lattice and the
  corresponding weights $w_\l$ for the Hermite quadrature.} 
  \label{tab:weights}  
\end{table}
The macroscopic fluid density $\rho$ and fluid velocity $\vec u$ are obtained from the moments of the distribution function,
\begin{align*}
	\sum_{\l=1}^{41} \sqrt g\, f_\l = \rho
	\quad,\quad
	\sum_{\l=1}^{41} \sqrt g\, \vec c_\l f_\l = \rho \vec u,
\end{align*}
where $\sqrt g$ is the square root of the determinant of the metric, originating from the integration measure on manifolds, $dV = \sqrt g\, dx^1 \cdots dx^D$. The density $\rho$ and
velocity $u^i$ fulfill the hydrodynamic conservation equations, which on a manifold read 
 \begin{align}
 	\label{eq:NS}
	\del_t \rho + \cov_i \left(\rho u^i \right) = 0, \qquad
	\del_t \left(\rho u^i \right) + \cov_j T^{ij} = 0,
\end{align} 
(covariant Navier-Stokes equations),
where $\cov$ denotes the covariant derivative (Levi-Civita connection) and
$T^{ij}$ is the energy-stress tensor. In the incompressible limit, the energy-stress tensor 
is given by 
\begin{align*}
	&T^{ij} = \Pi^{{\rm eq},ij} - \s^{ij}  \\
	&= \left( P\, g^{ij} + \rho\, u^i u^j \right)
	- \nu \left( \cov^j (\rho u^i) + \cov^i (\rho u^j)  + g^{ij} \cov_k (\rho u^k) \right),
\end{align*}
where in the first line $\Pi^{{\rm eq},ij}$ is the free momentum-flux tensor and $\s^{ij}$ denotes the viscous stress tensor. In the second line, $P = \rho\, \theta$ denotes the hydrostatic pressure, $\theta$ the normalized temperature, $\nu$ the kinematic viscosity, and $g^{ij}$ the components of the inverse metric tensor. (Note that the method is capable to handle very small but non-vanishing compressibilities, thus we keep terms like $\nabla_k (\rho u^k)$ in the equations although they are negligible.) We further work in the isothermal limit $\theta = 1$.

 The explicit discrete form of the equilibrium distribution is given by an expansion in tensor Hermite polynomials,
\begin{align*}
	f_\l^{\rm eq} &= \frac{w_\l}{\sqrt g} \bigg( a_{(0)}^{{\rm eq}} 
	+ \frac{1}{c_s^2} a_{(1)}^{{\rm eq},i} c_\l^i
	+ \frac{1}{2!\, c_s^4} a_{(2)}^{{\rm eq},ij} \Big( c_\l^i c_\l^j - c_s^2 \delta^{ij} \Big)\\
	&+ \frac{1}{3!\, c_s^6} a_{(3)}^{{\rm eq},ijk} \Big( c_\l^i c_\l^j
	c_\l^k - c_s^2 \big( \delta^{ij} c_\l^k + \delta^{jk}
	c_\l^i +   \delta^{ki} c_\l^j \big) \Big) \bigg),
\end{align*}
where the expansion coefficients read as follows:
\begin{align}
	\nonumber
	&a_{(0)}^{{\rm eq}} = \rho,\qquad
	a_{(1)}^{{\rm eq},i} = \rho u^i,\qquad 
	a_{(2)}^{{\rm eq},ij} = \rho \, c_s^2 \Delta^{ij} + \rho u^i u^j, \\
	\label{eq:hermite-coefficients-eq}
	&a_{(3)}^{{\rm eq},ijk} = \rho \,c_s^2 \left( \Delta^{ij} u^k + \Delta^{jk} u^i +
	\Delta^{ki} u^j \right) + \rho u^i u^j u^k.
\end{align}
Here, $w_\l$ are the lattice weights (see Table \ref{tab:weights}) and $\Delta^{ij} := g^{ij} - \delta^{ij}$ is a measure of the deviation from
flat space.

The forcing term $\mathcal F_\l$ is also expanded into Hermite polynomials, 
\begin{align}
	\nonumber
	\mathcal F_\l &= \frac{w_\l}{\sqrt g} \bigg( 
	\frac{1}{c_s^2} a_{(0)} F_\l^i c_\l^i
	+ \frac{1}{c_s^4} a_{(1)}^{i} F_\l^j \Big( c_\l^i c_\l^j - c_s^2 \delta^{ij} \Big)\\
	\label{eq:forcing_term}
	&+ \frac{1}{2 c_s^6} a_{(2)}^{ij} F_\l^k \Big( c_\l^i c_\l^j
	c_\l^k - c_s^2 \big( \delta^{ij} c_\l^k + \delta^{jk}
	c_\l^i +   \delta^{ki} c_\l^j \big) \Big) \bigg),
\end{align}
where $F_\l^i = - \Gamma^i_{jk} c_\l^j c_\l^k$ represents the inertial forces driving the fluid along the geodesics of the curved space, and $\Gamma^i_{jk}$ are the Christoffel symbols.
The expansion coefficients of the forcing term are given by
\begin{align*}
	&a_{(0)} = a_{(0)}^{{\rm eq}}, \qquad
	a_{(1)}^{i} = a_{(1)}^{{\rm eq},i},
	&a_{(2)}^{ij} = a_{(2)}^{{\rm eq},ij} - \s^{ij},
\end{align*}
where $\s^{ij} = -( 1 - \frac{1}{2\t}) \sum_\l \sqrt g \, c_\l^i c_\l^j \left( f_\l - f^{{\rm eq}}_\l \right)$ denotes the viscous stress tensor on the lattice.
External forces (needed to drive the fluid through the medium) are added to the lattice Boltzmann equation as described in Sec. \ref{sec:discrete_lattice_effects}.

From the metric tensor $g_{ij}$, the Christoffel symbols are calculated by
\begin{align*}
	\Gamma^i_{jk} = \frac 1 2 g^{im} \left( \del_k g_{jm} + \del_j g_{km} - \del_m g_{jk} \right),
\end{align*}
where the spatial derivatives of the metric can be computed very accurately by using isotropic lattice differentiation operators developed in Ref. \cite{thampi2013}, for example
\begin{align*}
	\del_k g_{jm}(\vec r) = \frac{1}{c_s^2 \dt} \sum_{\l=1}^{41} w_\l \,c_\l^k\, g_{jm}(\vec r + \vec c_\l \dt).
\end{align*}

Having all ingredients at hand for the LB Eq. 
(\ref{eq:LB}), the LB algorithm can be applied as usual: 
After assigning initial conditions to the macroscopic quantities $\rho$ and
$\vec u$, the distribution function $f$ is successively updated time step by 
time step according to the LB equation.

\subsection{Cancelation of discrete lattice effects}\label{sec:discrete_lattice_effects}

As already known from the lattice Boltzmann methods in flat Cartesian space \cite{Guo2002}, discrete lattice effects arise when a force is introduced into the lattice Boltzmann equation. These discrete lattice effects manifest themselves as spurious additional terms of order $\dt$ in the Navier-Stokes equations:
\begin{align}
	\label{eq:NS_discrete_lattice_effects}
	\del_t \rho + \cov_i \left(\rho u^i \right) 
	&= -\frac{\dt}{2} \left( \del_t A +  (\del_i - \G^j_{ij}) B^i \right) , \\
	\del_t \left(\rho u^i \right) + \cov_j T^{ij} 
	&=  -\frac{\dt}{2} \del_t B^i,
\end{align}
where $A := \sum_\l \sqrt g \, \FF_\l$ and $B^i := \sum_\l \sqrt g \, \FF_\l c_\l^i$ are the moments of the forcing term $\FF_\l$. As can be seen, the terms on the right-hand side of the equations act as spurious source terms and should thus be canceled. If the external force only depends on space and time (but not on the fluid velocity), this can be done through a simple redefinition of the fluid velocity (see Ref. \cite{Guo2002}). In our case, however, the inertial forces, $F_\l^i = - \Gamma^i_{jk} c_\l^j c_\l^k$, also depend on the microscopic fluid velocities $c_\l^i$, which affects the moments of the forcing term crucially. Moreover, our forcing term contributes not only to the momentum equation but also to the continuity equation, thus violating one of the necessary conditions on the forcing term considered in Ref. \cite{Guo2002}. For all these reasons, the spurious discrete lattice effects in our model cannot be canceled as straightforwardly as in the simple case considered in Ref. \cite{Guo2002}. However, it is still possible to eliminate spurious terms of order $\dt$ by correcting not only the velocity $\vec u$ but also the density $\r$ at each time step:
\begin{align}
	\label{eq:CE-summary1}
	\r 	&\quad\longrightarrow\quad \text{\leer{$\uu^i$}} \rr = \text{\leer{$u^i$}} \r + \dt\, R(\rr,\uu), \\
	\label{eq:CE-summary2}
	\r u^i &\quad\longrightarrow\quad \rr \uu^i = \r u^i + \dt\, U^i(\rr,\uu),
\end{align}
where $R(\rr,\uu)$ and $U^i(\rr,\uu)$ are correction terms, given by	
\begin{align*}	
	R(\rr,\uu) &= - \frac{1}{2} \left( \G^i_{ij} \rr  \uu^j + \G^i_{ji} \rr \uu^j \right), \\	
	U^i(\rr,\uu) &= - \frac{1}{2} \left( \G^i_{jk} \Pi^{{\rm eq},jk} + \G^j_{jk} \Pi^{{\rm eq},ki}  + \G^j_{kj} \Pi^{{\rm eq},ki} \right),
\end{align*}
where $\Pi^{{\rm eq},ij} = \rr\, c_s^2  g^{ij} + \rr \uu^i \uu^j$  denotes the free momentum-flux tensor. The correction terms $R$ and $U^i$ correspond to the moments of the forcing term and account for inertial effects on the fluid velocity and on the energy-stress tensor, which  originate from the covariant derivatives in the Navier-Stokes Eqs. (\ref{eq:NS}).
Equations (\ref{eq:CE-summary1}) and (\ref{eq:CE-summary2}) define a system of coupled quadratic equations in $\uu$, which can be solved numerically for $\rr$ and $\uu$ (e.g. using Newton's algorithm, which converges to the solution rapidly after a few iterations; see Appendix \ref{sec:newton_algorithm}).
Additionally, the forcing term Eq. (\ref{eq:forcing_term}) is rescaled as follows:
\begin{align}\label{eq:CE-summary3}
	\FF_\l &\quad\longrightarrow\quad \left( 1 - \frac{1}{2\t}\right) \FF_\l.
\end{align}
  The equilibrium distribution as well as the forcing term are evaluated with the corrected density and velocity at each time step, $f^{{\rm eq}} = f^{{\rm eq}}(\rr,\uu)$, $\FF = \FF(\rr,\uu)$. 
An additional external force $\vec F^{\rm{ext}}$ can be added straightforwardly to the lattice Boltzmann equation, replacing $F_\l^i$ by $(F_\l^i + F_{\rm
ext}^i)$ in the forcing term Eq. (\ref{eq:forcing_term}), which also results in an additional term in the correction function for the macroscopic velocity:
\begin{align*}	
	U^i(\rr,\uu) &\quad\longrightarrow\quad U^i(\rr,\uu) + \frac{1}{2} \rr  F_{\rm
ext}^i.
\end{align*}


As can be shown rigorously by a Chapman-Enskog expansion (see Appendix \ref{sec:app} for the details), these corrections lead to the correct macroscopic Eqs. (\ref{eq:NS}), where spurious terms of order $\dt$ are canceled. In Sec. \ref{sec:discrete-lattice-effects}, we show that by using the corrected density Eq. (\ref{eq:CE-summary1}) and velocity Eq. (\ref{eq:CE-summary2}) as well as the modified forcing term Eq. (\ref{eq:CE-summary3}), the method is improved considerably. In the following, we will drop the $\,\hat{}\,$ symbol for the corrected density and velocity for simplicity.

\section{Poiseuille Flow in Curved Space}\label{sec:campylotic-media}

In this section, we study the Poiseuille-like channel flow in curved space by 
investigating flow through different types of curved space. To this end, we disturb the flat space by adding local perturbations to the metric tensor, resulting in a physical curvature of space. This has to be distinguished from a simple coordinate transformation into curvilinear coordinates, which would not introduce spatial curvature. 
In the following, we will work in Cartesian-like coordinates $(x^1,x^2,x^3) = (x,y,z)$, such that the metric tensor of the flat space reads $g_0 = \mathbbm 1$. The perturbed metric tensor with $N$ perturbations (labeled by an index $i$) is then given by 
\begin{align}\label{eq:metric}
	g 
	= (1 + \dg) \cdot \mathbbm{1}
	= \left( 1 + \sum_{i=1}^N \dg_i \right) \cdot \mathbbm{1},
\end{align}
where $\dg_i$ denotes the contribution of the $i$th perturbation. The individual metric perturbations are chosen to be of the shape
\begin{align}\label{eq:radial-campylons}
	\dg_i = \begin{cases} 
				-a_0 \cos^2\left( \pi \frac{r_i}{r_0} \right), & r_i \leq \frac{r_0}{2}, \\
				0, & r_i > \frac{r_0}{2},
			\end{cases}	
\end{align}
where $r_i = \| \vec r - \vec r_i \|$ denotes the distance to the center $\vec r_i$ of the $i$-th perturbation. The parameters $a_0$ and $r_0$ characterize the amplitude and the range of the perturbation, respectively. We further define the density $n$ of the perturbations by $n = N/{V_0}$, where $V_0$ denotes the volume of the flat background medium. Since the individual perturbations possess a rotational symmetry, we will call this type ``radial perturbations.''
In the simulations, we model a channel with periodic boundary conditions in the streamwise direction (at $x=0, x=L_x$) and fixed walls at the perpendicular boundaries (at $y=0, y=L_y$ and $z=0, z=L_z$). At the walls, we impose zero-velocity boundary conditions on the fluid, which enters the algorithm through the equilibrium distribution $f^{\rm eq}$ being evaluated with $u = 0$ at the wall nodes. The fluid is driven through the medium by a constant external force corresponding to a pressure drop of $\nabla P = 10^{-5}$ in $x$ direction. For all the two-dimensional simulations, we use a $D3Q41$ lattice of size $L_x \times L_y \times L_z = 256 \times 256 \times 1$, and set the discretization step to $\dt = \dx^i = 128/L_x$. For the viscosity of the fluid, we choose $\nu = c_s^2 (\tau - \frac{1}{2})\, \dt = c_s^2/4$, which fixes the relaxation parameter $\tau$ correspondingly. The three-dimensional simulations, on the other hand, are performed on a lattice of size $L_x \times L_y \times L_z = 64 \times 64 \times 64$ with $\dt = \dx = \dy = \dz = 1$ and $\nu = c_s^2 (\tau - \frac{1}{2})\, \dt = c_s^2/2$.

After the fluid has reached the stationary state, we measure the conserved flux of the flow, which in curved space is given by
\begin{align*}
	\Phi = \int_S g(\rho\, \vec u, \vec n_x)\, dS 
	= \int_S \rho\, u^x \sqrt g \, dy\, dz,
\end{align*}
where $\vec n_x = \vec e_x/\sqrt{g_{xx}}$ denotes the unit vector in $x$ direction and the integral runs over a channel cross-section with area element $dS = \sqrt{g_{yy}\,g_{zz}}\, dy\, dz $ and total area $S = \int_S dS$.
Since we want to study the transport properties independently of the specific size of the  medium, we consider the spatial average of the flux, given by
\begin{align}\label{eq:flux-average}
	\langle \Phi \rangle 
	= \frac{1}{S} \int_S g(\rho\, \vec u, \vec n_x)\, dS
	= \frac{\Phi}{S}.
\end{align}
In particular, we are interested in the relative deviation of the flux from the corresponding flux of the standard Poiseuille flow in flat space, $\Phi_0$, which is described by
\begin{align*}
	\Delta \Phi := \frac{\langle \Phi_0 \rangle - \langle \Phi \rangle}{\langle \Phi_0 \rangle}.
\end{align*}

\subsection{2D curved media}\label{sec:campylotic-media-regular}

\begin{figure}[ht]
	\includegraphics[width=.7\columnwidth]{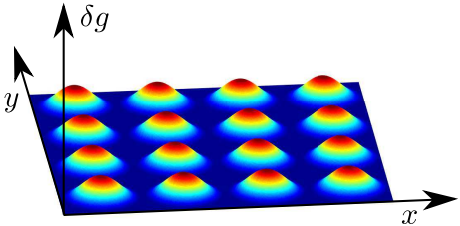}
	\caption{Curved medium with regularly arranged metric perturbations. The colors illustrate the strength of the perturbations, ranging from $0$ (blue) to $a_0$ (red).}
	\label{fig:metric_regular}
\end{figure}

For simplicity we start with a two-dimensional periodic arrangement of metric perturbations, where the positions $\vec r_n$ of the perturbations are confined to a regular square lattice,
\begin{align*}
	\vec r_n 
	= \vec r_{n_x,n_y} 
	= (n_x, n_y) \cdot \l, \qquad n_x,n_y \in \Z,
\end{align*} 
where $\l = \sqrt{V_0/N}$ denotes the characteristic distance between the perturbations and $V_0 = L_x L_y \dx \dy$ is the volume of the flat background space (see Fig. \ref{fig:metric_regular} for a sketch).
\begin{figure*}[ht]
\subfigure[]{
\includegraphics[width=0.31\textwidth]{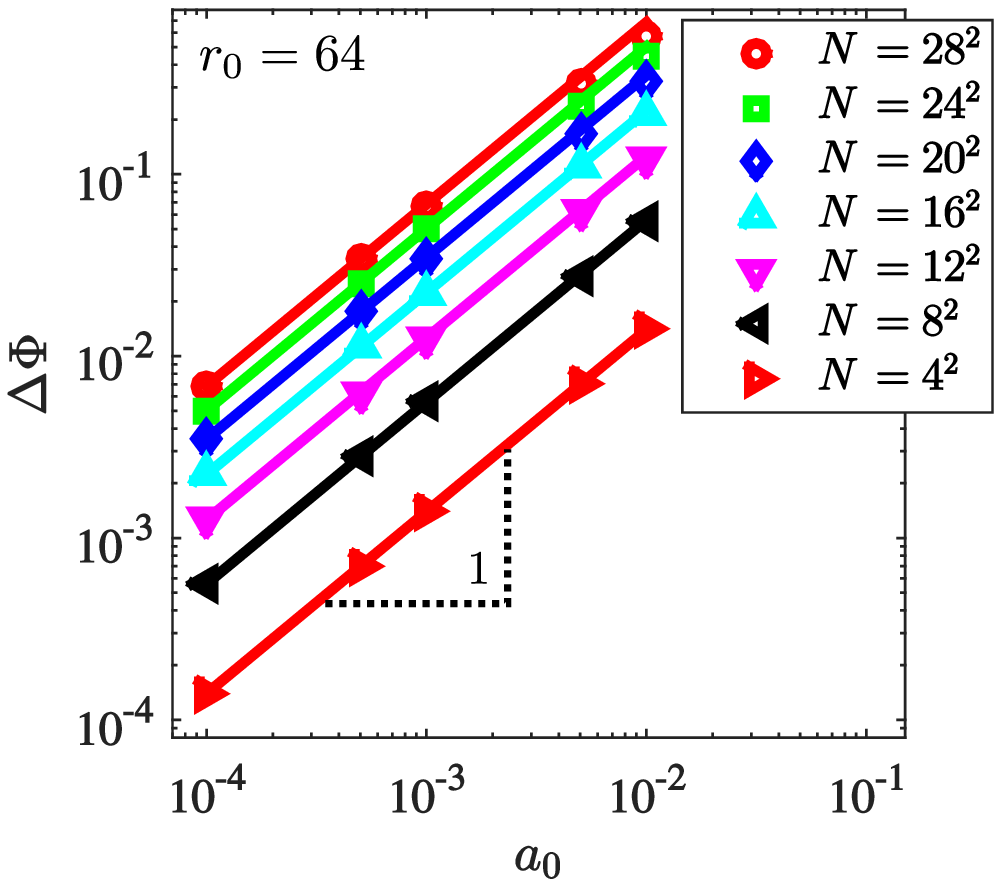}
\label{fig:paper_flux_regular_a0}
}
\hspace*{\fill} 
\subfigure[]{
\includegraphics[width=0.31\textwidth]{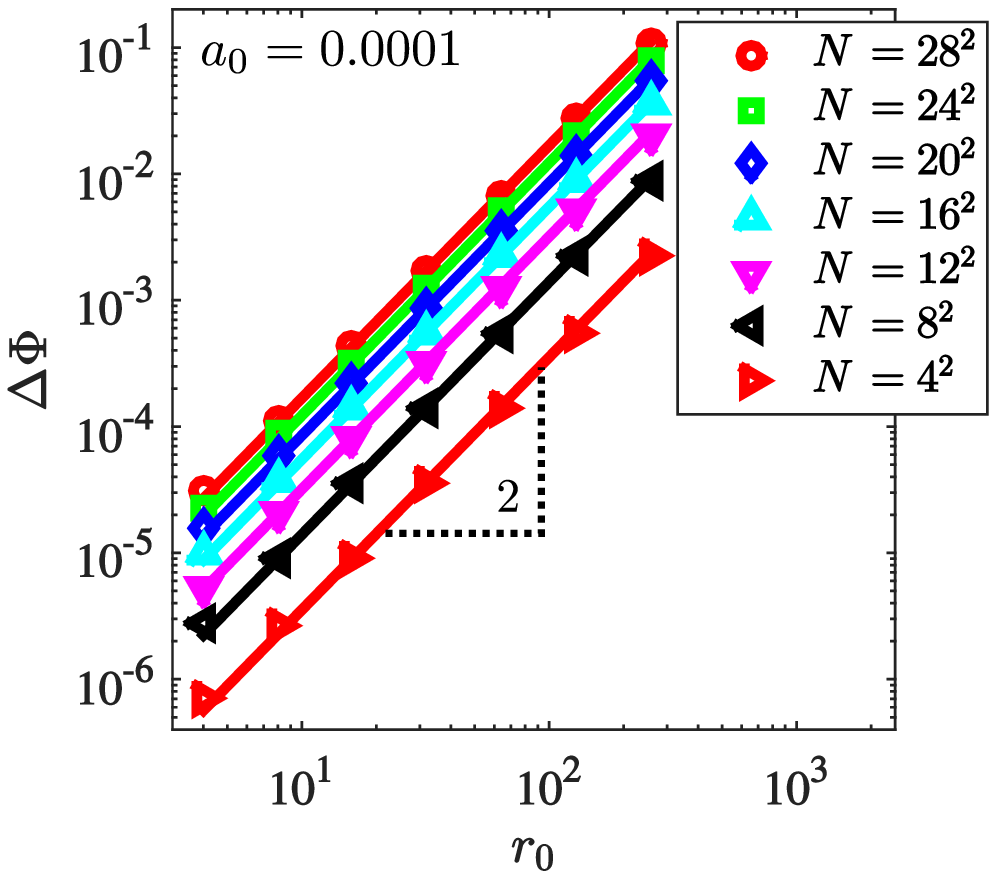}
\label{fig:paper_flux_regular_r0}
}
\hspace*{\fill} 
\subfigure[]{
\includegraphics[width=0.31\textwidth]{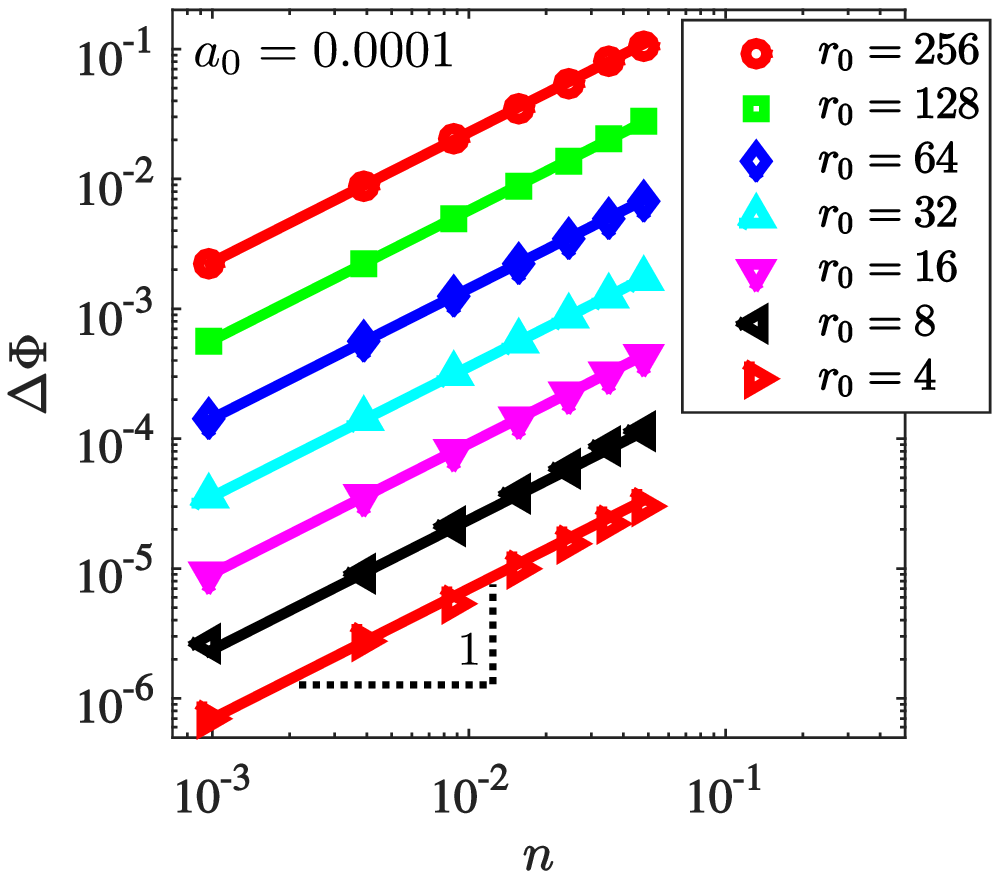}
\label{fig:paper_flux_regular_n}
}
\caption{Flux deficit $\Delta \Phi = \frac{\langle \Phi_0 \rangle - \langle \Phi \rangle}{\langle \Phi_0 \rangle}$ as function of the metric perturbation amplitude $a_0$ (a), range $r_0$ (b), and density $n$ (c) for a 2D simulation with $N$ \textit{regularly} arranged metric perturbations.} 
\label{fig:paper_flux_regular}
\end{figure*} 
Because of the periodicity of the metric perturbations, one might be inclined to simulate only one perturbation with periodic boundary conditions in all directions. However, since we put walls at $y=0, y=L_y$ and since we aim at extending this system to a random arrangement of perturbations, we simulate the entire regular system. 
In order to find the dependence of the flux on the perturbation amplitude $a_0$, range $r_0$, and density $n = N/{V_0}$, we have performed simulations in a wide parameter range. Our simulations range from dilute media ($r_0 < \l$), for which the metric perturbations are spatially separated from each other, to very dense media ($r_0 \gg \l$), where the metric perturbations overlap and form large clusters.
As soon as the fluid reaches its equilibrium state, the average flux is measured according to Eq. (\ref{eq:flux-average}). In Fig. \ref{fig:paper_flux_regular} the flux deficit $\Delta \Phi$ is plotted against the metric parameters in a double-logarithmic plot. As can be seen, all the curves fulfill a power law with integer slopes, which is even valid in the regime of overlapping perturbations, $r_0 > \l$. From the power law behavior, we can infer the following functional dependence of $\Delta \Phi$ on the metric parameters:
\begin{align}\label{eq:delta_phi}
	\Delta\Phi = C_0 \,a_0 \, r_0^2 \, n + \OO(a_0^2),
\end{align}
where $C_0$ is a constant. We recognize that this expression resembles the average metric perturbation, which we define as
\begin{align}
	\label{eq:def-dg}
	\langle \dg \rangle 
	= \frac{1}{V_0} \int \dg\ d^Dx
\end{align} 
where $d^Dx = dx^1 \cdots dx^D$. Plugging in the explicit expression for the radial  perturbations (\ref{eq:radial-campylons}) in two dimensions, we obtain
\begin{align}
	\label{eq:dg-regular}
	\langle \dg \rangle &=  -\frac{\pi^2-4}{8\pi}\, a_0\, r_0^2\, n.	
\end{align}
Thus, we can rewrite Eq. (\ref{eq:delta_phi}) as follows:
\begin{align}\label{eq:flux-first-order}
	\langle \Phi \rangle = \langle \Phi_0 \rangle \cdot \left( 1 + C_1 \cdot \langle \dg \rangle + \OO(a_0^2) \right),
\end{align}
where $C_1 = 1.500 \pm 0.003$.
This relation describes the main contribution of the metric parameters to the flux. In order to check that relation Eq. (\ref{eq:flux-first-order}) is not just a coincidence which only holds for this special type of radial perturbations, we have also studied metric perturbations of a different shape, given by the metric function	
\begin{align}\label{eq:cubic-campylons}
	\d\widetilde{g}_i = \begin{cases} 
				-a_0 \cos^2\left( \pi \frac{|x-x_i|}{r_0} \right) \cos^2\left( \pi \frac{|y-y_i|}{r_0} \right), \\
				 \qquad\qquad\qquad\qquad
				 \text{if } |x-x_i|\leq \frac{r_0}{2},|y-y_i| \leq \frac{r_0}{2}, \\
				0, 
				\qquad\qquad\qquad\quad
				\text{else},
			\end{cases}	
\end{align}
where $\vec r_i = (x_i,y_i)$ denotes the center of the $i$th perturbation. We will refer to this type as ``square perturbations,'' as their support is a square, and mark all related quantities with a tilde symbol.
For the square perturbations, the average metric perturbation is given by
\begin{align*}
	\langle \d\widetilde{g} \rangle
	=  -\frac{1}{4}\, a_0\, r_0^2\, n,	
\end{align*}
and only differs from the corresponding expression for the radial perturbations Eq. (\ref{eq:dg-regular}) in the constant prefactor. 

\begin{figure}[ht]
\includegraphics[width=0.9\columnwidth]{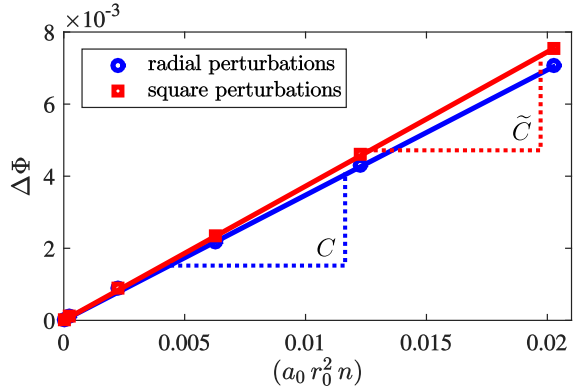}
\caption{Flux deficit $\Delta \Phi = \frac{\langle \Phi_0 \rangle - \langle \Phi \rangle}{\langle \Phi_0 \rangle}$ as function of the product $a_0\, r_0^2\, n$ for two different types of metric perturbations. $a_0 = 0.001$ and $r_0 = 16$ are kept fixed while the number of perturbations is varied.}
\label{fig:paper_flux_radial_and_cubic}
\end{figure} 
Figure \ref{fig:paper_flux_radial_and_cubic} shows the flux deficit $\Delta \Phi$ as function of the parameter combination $(a_0\, r_0^2\, n)$ for the two different types of metric perturbations. It can be seen that the flux in the simulations with square  perturbations exhibits the same functional dependence on the metric parameters,
\begin{align*}
	\Delta \widetilde\Phi = {\widetilde C}_0\, a_0\, r_0^2\, n + \OO(a_0^2),
\end{align*}
with a different slope ${\widetilde C}_0$. From the linear fits we obtain $ C_0 = 0.350 \pm 0.004$ for the slope of the radial perturbations and ${\widetilde C}_0 = 0.374 \pm 0.003$ for the square perturbations. Assuming that relation Eq. (\ref{eq:flux-first-order}) represents a universal flux law holding for both types of perturbations, the ratio between $\Delta \Phi$ and $\Delta \widetilde \Phi$ has to be equal to the ratio between $\langle \dg \rangle$ and $\langle \d\widetilde{g} \rangle$. We find:
\begin{align*}
	\frac{\Delta \widetilde \Phi}{\Delta \Phi} 
	&= \frac{{\widetilde C}_0}{C_0} = 1.069 \pm 0.004,  \\
	\frac{\langle \d\widetilde{g} \rangle}{\langle \dg \rangle} 
	&= \frac{2 \pi}{\pi^2 - 4} = 1.070.
\end{align*}
Thus, we conclude that --- at least for the two types of metric perturbations considered here --- relation Eq. (\ref{eq:flux-first-order}) holds independently of the specific shape of the perturbations.

As already mentioned, relation Eq. (\ref{eq:flux-first-order}) is expected to be only a linear approximation of the real functional dependence of $\Phi$ on $\langle \dg \rangle$. In order to find the higher-order contributions, we have also performed simulations for stronger spatial deformations, where non-linear effects are expected to appear. To this end, we have again varied all the metric parameters within a wide range of perturbation amplitudes $a_0$ and also included the case of negative amplitudes $a_0 < 0$. 
\begin{figure}[ht]
\includegraphics[width=1\columnwidth]{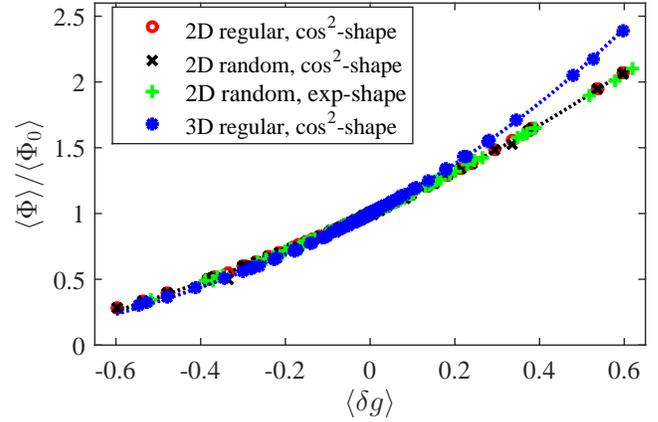}
\caption{Flux ratio $\langle \Phi \rangle / \langle \Phi_0 \rangle$ as function of the average strength of the metric perturbations $\langle \dg \rangle$ for different types of curved media (2D regular, 2D random and 3D regular with $\cos$- or $\exp$-shaped metric perturbations). For each medium type, all data points collapse onto a single common curve.}
\label{fig:paper_flux_nonlinear}
\end{figure} 
Figure \ref{fig:paper_flux_nonlinear} shows the resulting curve of $\langle \Phi \rangle / \langle \Phi_0 \rangle$ as a function of $\langle \dg \rangle$. As can be seen, all the points fall on one single curve, which leads to the conclusion that even beyond the first-order approximation in $a_0$, the flux is well-described as a function of $\langle \dg \rangle$. From Fig. \ref{fig:paper_flux_nonlinear}, we find the following flux law:
\begin{align}\label{eq:flux-law-2D}
	\langle \Phi \rangle = \langle \Phi_0 \rangle \cdot 
	\left( 1 + \a^2 \cdot \langle \dg \rangle 
	 + 2 \a^2 \b \cdot \langle \dg \rangle^2 \right),
\end{align}
where $\a = 1.223 \pm 0.001$ and $\b = 0.165 \pm 0.001$ are fitting parameters. This flux law is valid for metric perturbations $\langle \dg \rangle = \OO(1)$ and supports our expectations about transport in curved media: By adding a metric perturbation $\dg$ to the flat space, the space becomes either stretched ($\dg > 0$) or compressed ($\dg < 0$). 
In the first case, $\dg > 0$, the cross-section of the medium is increased by the presence of the metric perturbations, such that the transport through the medium is enhanced and the flux increases compared to the flat space.
For negative metric perturbations, $\dg < 0$, on the other hand, the channel cross-section is decreased, and the fluid transport deteriorates compared to the flat space.

\begin{figure}[ht]
	\includegraphics[width=.7\columnwidth]{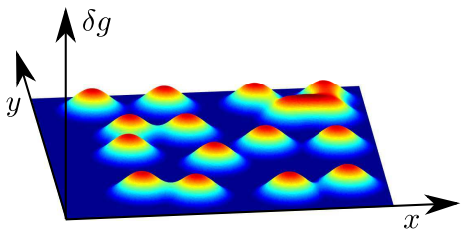}
	\caption{Curved medium with randomly arranged metric perturbations. The colors illustrate the strength of the metric perturbation, ranging from $0$ (blue) to $a_0$ (red).}
	\label{fig:metric_random}
\end{figure} 
 
We also have investigated the case of randomly distributed metric perturbations (see Fig. \ref{fig:metric_random} for a sketch) in order to study the dependence of the Poiseuille flow on the spatial order of the metric perturbations.
For small amplitudes $a_0$, the spatial arrangement of perturbations does not affect the linear order of the average metric perturbation, given by Eq. (\ref{eq:dg-regular}). Indeed, we find that the dependence of $\Delta \Phi$ on the metric parameters is exactly the same as in the regular case, given by Eq. (\ref{eq:delta_phi}), and the flux is thus well-described as a function of $\langle \dg \rangle$.
Even for larger amplitudes $a_0$, the flux curve does not change notably from the case of regularly arranged metric perturbations, as can be seen in Fig. \ref{fig:paper_flux_nonlinear}. We find that the flux follows Eq. (\ref{eq:flux-law-2D}) with fitting parameters $\a = 1.224 \pm 0.002$ and $\b = 0.158 \pm 0.005$, in agreement with the corresponding values for regularly arranged metric perturbations.
 Thus, we conclude that --- within the range of parameters studied --- the spatial order of the metric perturbations does not affect the behavior of the flux significantly.

In order to check that the flux law also holds for metric perturbations with non-compact support, we have also performed simulations for exponentially shaped perturbations of the form $\dg_i = -a_0 \exp\left( -r_i/r_0 \right)$, distributed randomly in space. For this  perturbation type, the average metric perturbation yields $\langle \dg \rangle =  -2 \pi \, a_0\, r_0^2\, n$,
and the simulations results have been added to Fig. \ref{fig:paper_flux_nonlinear} with fitting coefficients $\a = 1.225 \pm 0.001$ and $\b = 0.144 \pm 0.002$. As can be seen, the flux curve coincides with the other flux curves in 2D, which supports our claim that the flux law Eq. (\ref{eq:flux-law-2D}) holds universally.

\begin{figure}[ht]
	\includegraphics[width=.7\columnwidth]{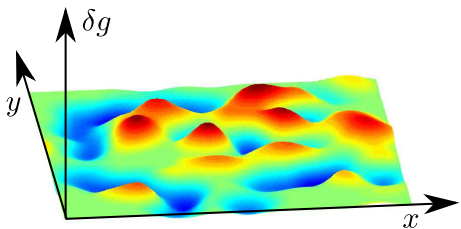}
	\caption{Curved medium with mixed positive and negative metric perturbations. The colors illustrate the strength of the metric perturbation, ranging from $-a_0$ (blue) to $+a_0$ (red).}
	\label{fig:paper_metric_mixed_sign}
\end{figure} 

Additionally, we have also simulated media with both positive and negative metric perturbations (see Fig. \ref{fig:paper_metric_mixed_sign} for a sketch). If the flux law Eq. (\ref{eq:flux-law-2D}) holds in general, any non-trivial configuration of perturbations with $\langle \dg \rangle = 0$ should give the same value for the flux, namely $\langle \Phi \rangle = \langle \Phi_0 \rangle$. To test this statement, we have performed simulations for a wide range of metric parameters, $a_0 \in \{\pm 0.0001,\pm 0.001,\pm 0.01\}$, $r_0 \in \{4,8,16,32,64,128\}$, $N \in \{16,64\}$, all satisfying $\langle \dg \rangle = 0$. Indeed, we find that all simulations give the same flux $\Phi_0$ with a standard deviation of about $0.2 \%$.

\subsection{3D curved media}\label{sec:campylotic-media-regular3D}

\begin{figure*}[ht]
\subfigure[]{
\includegraphics[width=0.31\textwidth]{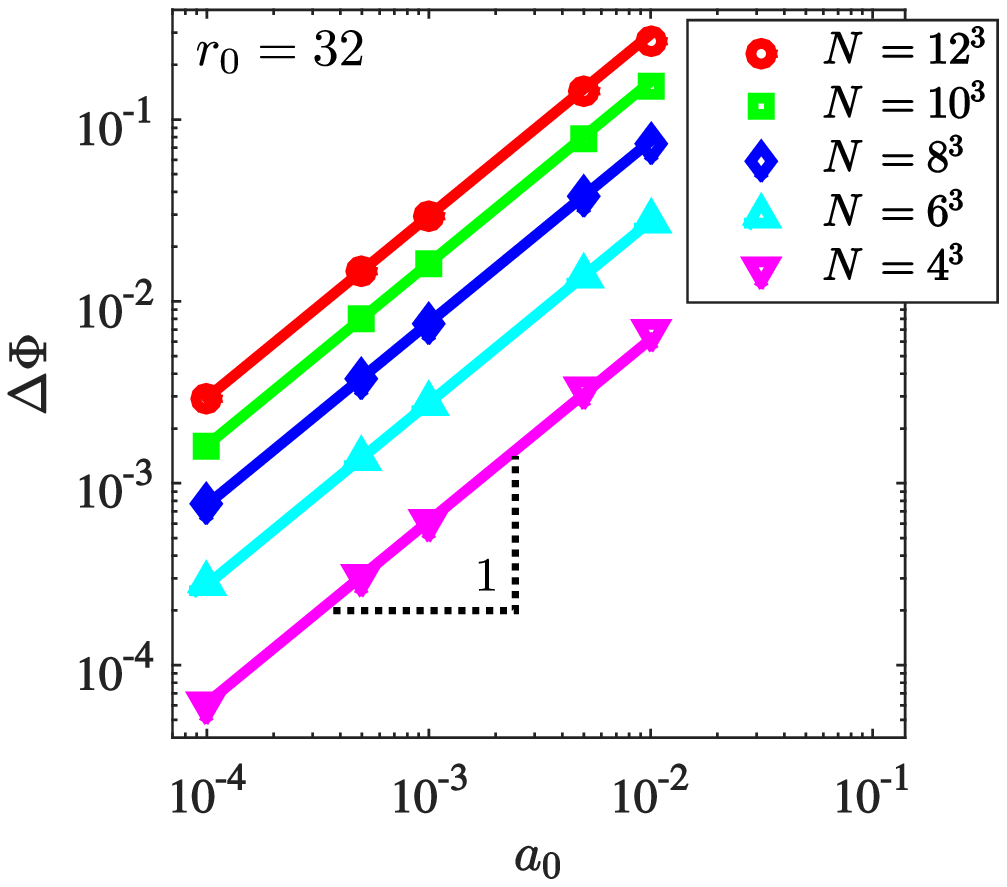}
\label{fig:paper_flux_regular3D_a0}
}
\hspace*{\fill} 
\subfigure[]{
\includegraphics[width=0.31\textwidth]{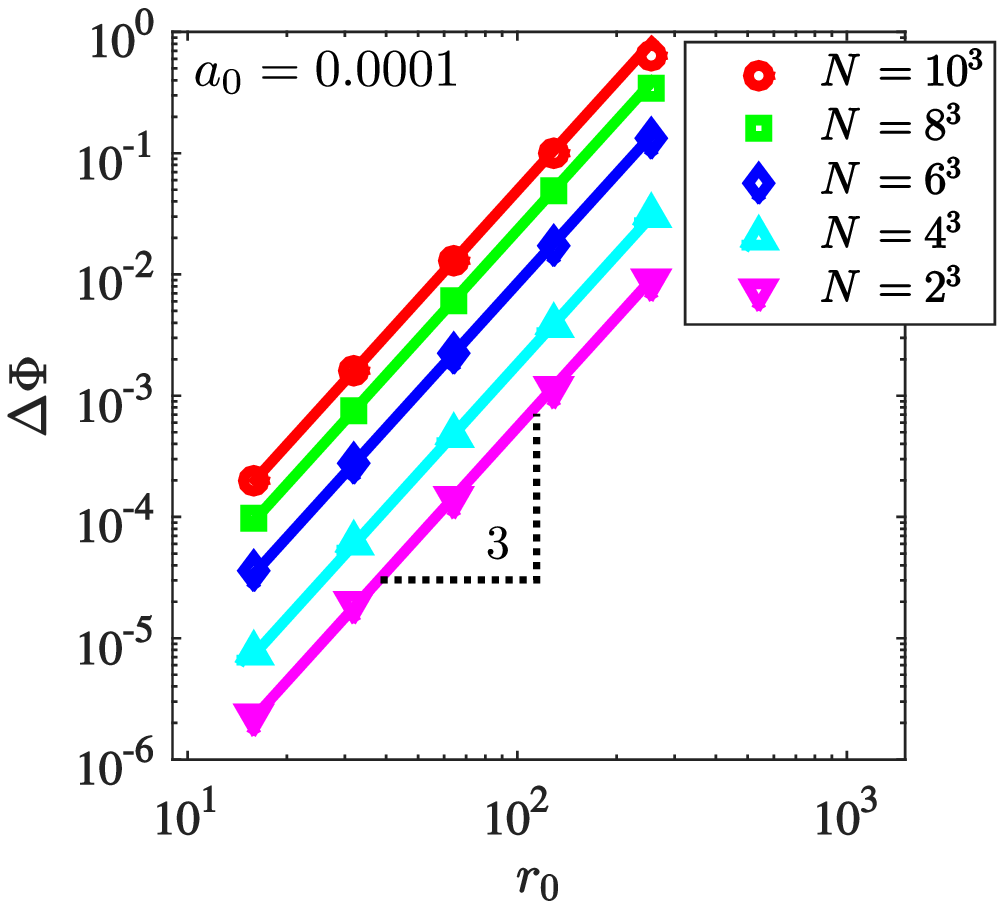}
\label{fig:paper_flux_regular3D_r0}
}
\hspace*{\fill} 
\subfigure[]{
\includegraphics[width=0.31\textwidth]{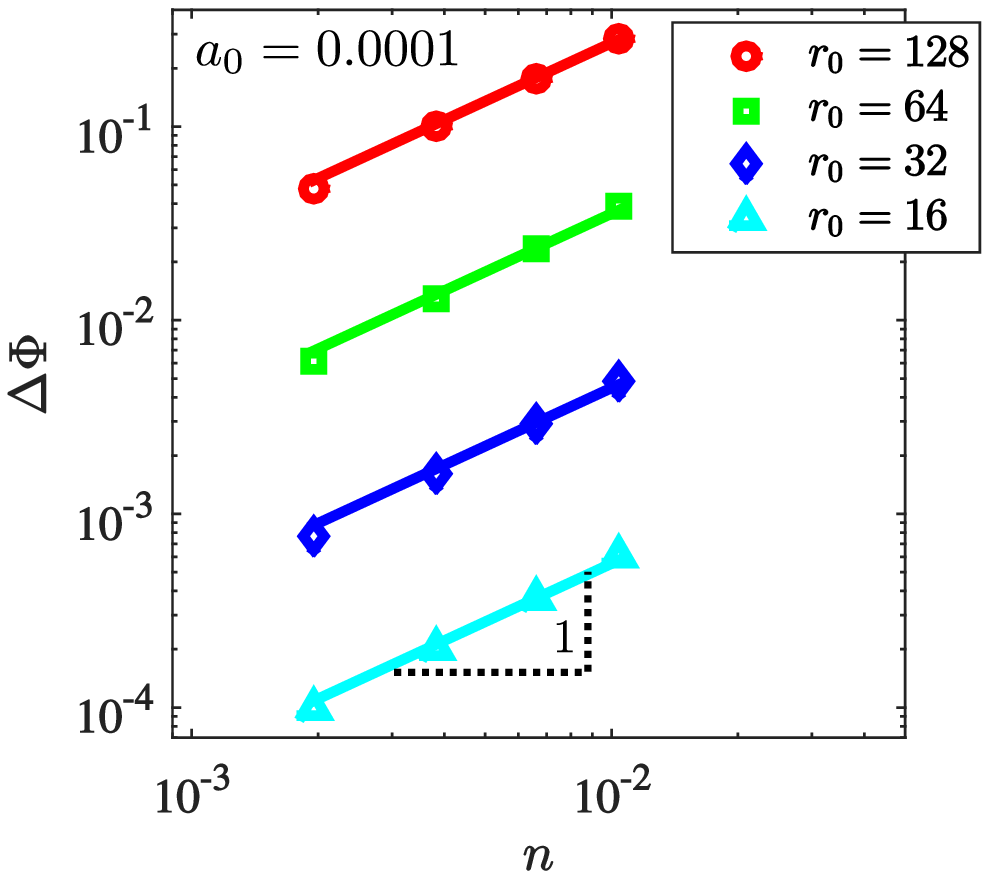}
\label{fig:paper_flux_regular3D_n}
}
\caption{Flux deficit $\Delta \Phi = \frac{\langle \Phi_0 \rangle - \langle \Phi \rangle}{\langle \Phi_0 \rangle}$ as function of the metric perturbation amplitude $a_0$ (a), range $r_0$ (b), and density $n$ (c) for a 3D simulation with \textit{regularly} arranged metric perturbations.} 
\label{fig:paper_flux_regular3D}
\end{figure*} 

Next, we consider a three-dimensional curved channel equipped with $N$ regularly arranged metric perturbations at positions
\begin{align*}
	\vec r_n 
	= \vec r_{n_x,n_y,n_z} 
	= (n_x, n_y, n_z) \cdot \l, \qquad n_x,n_y,n_z \in \Z,
\end{align*} 
where $\l = \sqrt[3]{V_0/N}$ denotes the characteristic distance between the metric perturbations and $V_0 = L_x L_y L_z \dx \dy \dz$ is the volume of the flat background space. 

Like for the two-dimensional cases, we have studied the dependence of the Poiseuille flow on the metric parameters as depicted in Fig. \ref{fig:paper_flux_regular3D}. We find that --- at leading order in the metric perturbations --- the flux deficit is given by
\begin{align*}
	\Delta \Phi = C_0 \,a_0 \, r_0^3 \, n + \OO(a_0^2),
\end{align*}
$C_0$ being a constant. In analogy to the two-dimensional case, the expression $(a_0 \, r_0^3 \, n)$ corresponds to the average metric perturbation in three dimensions, given by
\begin{align*}
	\langle \dg \rangle = -\frac{\pi^2-6}{12\pi}\, a_0\, r_0^3\, n.
\end{align*} 
Plotting the flux versus $\langle \dg \rangle$ for $|\langle \dg \rangle| \leq 0.6$, as shown in Fig. \ref{fig:paper_flux_nonlinear}, we again find that even beyond the leading order, all points fall on one single curve, which can be described by the flux law
\begin{align}\label{eq:flux-law-3D}
	\langle \Phi \rangle = \langle \Phi_0 \rangle \cdot 
	\left( 1 + \a^3 \cdot \langle \dg \rangle 
	 + 3 \a^3 \b \cdot \langle \dg \rangle^2 \right),
\end{align}
where the fitting parameters take the values $\a = 1.21 \pm 0.01$ and $\b = 0.17 \pm 0.01$ and thus agree with the corresponding values of the two-dimensional flux law Eq.  (\ref{eq:flux-law-2D}).
This suggests that the flux law might be generalized as follows:
\begin{align}\label{eq:flux-law}
	\langle \Phi \rangle = \langle \Phi_0 \rangle \cdot 
	\left( 1 + \a^D \cdot \langle \dg \rangle 
	 + D \a^D \b \cdot \langle \dg \rangle^2 \right),
\end{align}
where $D$ denotes the dimension of the system and universal constants $\a \approx 1.22$ and $\b \approx 0.16$.

\section{Comparison with Previous Study}

The findings presented in this paper generalize the results of the previous study \cite{mendoza2013} of three-dimensional curved media equipped with randomly distributed  metric perturbations of exponential shape, $\dg_i = -a_0 \exp\left( -r_i/r_0 \right)$. In the previous study, only positive amplitudes (corresponding to negative values of $\langle \dg \rangle$) were considered, and the flux was found to obey the laws
\begin{align}
	\label{eq:flux-laws1}
	\Phi_0 - \Phi &= A_0 \frac{N/N_0}{1+(N/N_0)^2},  \\
	\label{eq:flux-laws2}
	\Phi_0 - \Phi &= A_0 \frac{(\epsilon/\epsilon_0)^3}{1+(\epsilon/\epsilon_0)^6},
\end{align}
where $N$ denotes the number of metric perturbations, $\epsilon= r_0/\l$ is the dimensionless  deformation, $\l = \sqrt[3]{V_0/N}$ is the characteristic length, $V_0$ denotes the volume of the flat background space, and $A_0, N_0$, and $\epsilon_0$ are fitting parameters. Calculating the average metric perturbation for this perturbation type, we obtain
\begin{align*}
	\langle \dg \rangle &=  -8\pi\, a_0\, r_0^3\, n
	= -8\pi\, a_0\, \epsilon^3.
\end{align*}
With that, Eqs. (\ref{eq:flux-laws1}) and (\ref{eq:flux-laws2}) can be combined into a single equation of the following form:
\begin{align}
	\label{eq:flux-laws12}
	\frac{\langle \Phi \rangle}{\langle \Phi_0 \rangle} 
	\approx 1 + \widetilde{A}_0 \frac{\langle \dg \rangle / \langle \dg \rangle_0}{1+(\langle \dg \rangle  / \langle \dg \rangle_0)^2},
\end{align}
where $\widetilde{A}_0$ and $\langle \dg \rangle_0$ are constants.
Here, we have switched to the spatially averaged flux, dividing by the cross-sectional area $S$ according to Eq. (\ref{eq:flux-average}). 

Restricting to positive amplitudes, i.e., $\langle \dg \rangle < 0$, the flux curves depicted in Fig. \ref{fig:paper_flux_nonlinear} are indeed well-described by  Eq. (\ref{eq:flux-laws12}), thus showing that the flux law in Ref. \cite{mendoza2013} is consistent with our present results. 
However, Eq. (\ref{eq:flux-laws12}) fails to describe the correct behavior for negative amplitudes (corresponding to $\langle \dg \rangle > 0$) and should thus be replaced by the more general flux law given by Eq. (\ref{eq:flux-law}). This flux law also explains the power law dependence of the flux deficit on $\epsilon$ for small $a_0$ found in Ref. \cite{mendoza2013}, since at leading order in $a_0$ we have $(\langle \Phi_0 \rangle - \langle \Phi \rangle) \sim \langle \dg \rangle \sim r_0^3\, n = \epsilon^3$.

\section{Method Validation}\label{sec:method-validation}

\subsection{Improvement by Cancelation of Discrete Lattice Effects}\label{sec:discrete-lattice-effects}

\begin{figure}[ht]
\subfigure[]{
\includegraphics[width=0.46\columnwidth]{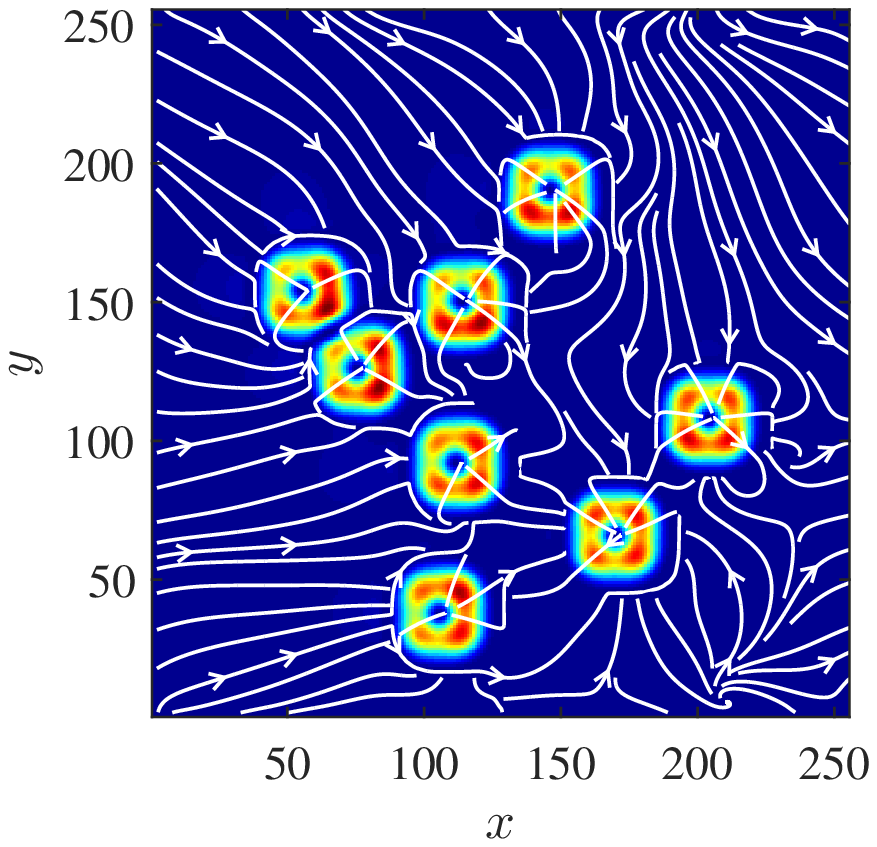}
\label{fig:paper_corrections_streamlines1}
}
\hspace*{\fill} 
\subfigure[]{
\includegraphics[width=0.46\columnwidth]{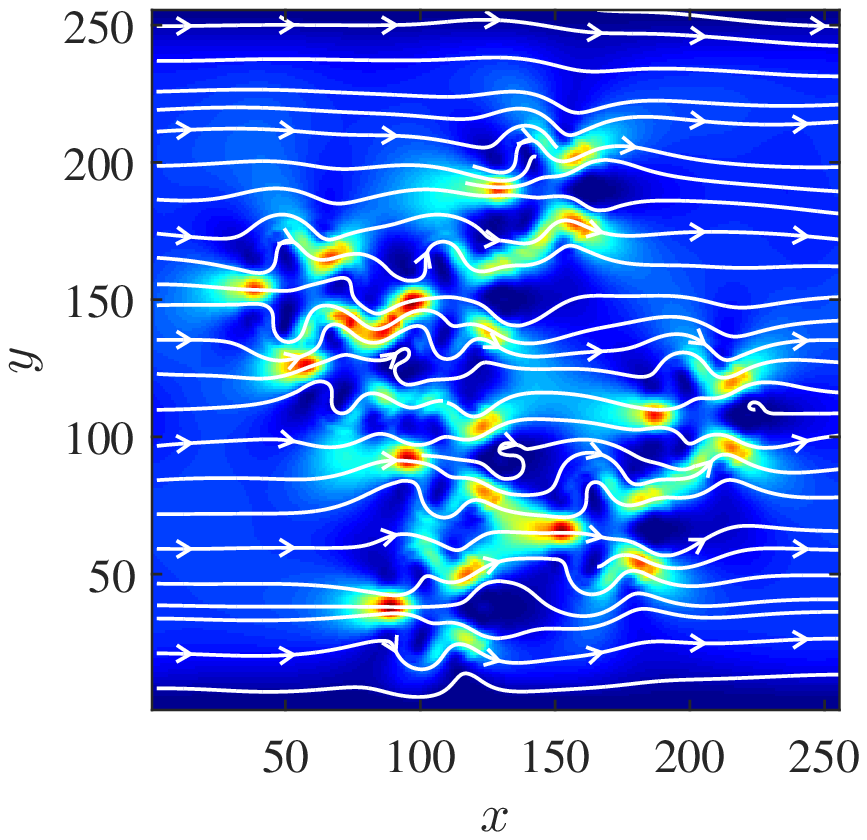}
\label{fig:paper_corrections_streamlines2}
}
\caption{Velocity streamlines of the fluid for a simulation with $N=8$ metric perturbations with amplitude $a_0 = 0.1$ and range $r_0 = 22$. The colors illustrate the absolute value of the velocity. (a) Simulation with discrete lattice effects and spurious source terms. (b) Corrected model.} 
\label{fig:paper_corrections_streamlines}
\end{figure}

\begin{figure}[ht]
\includegraphics[width=.9\columnwidth]{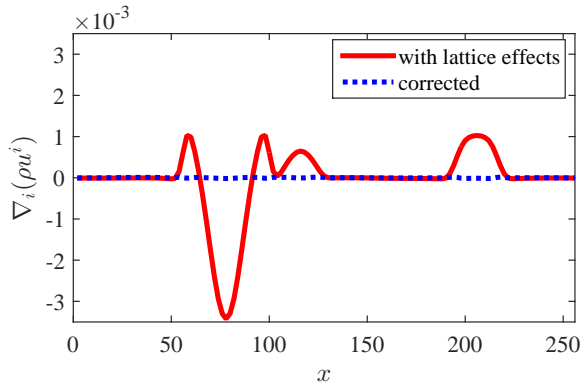}
\caption{The divergence field $\nabla_i (\rho u^i)$ as a function of $x$ at the cross-section $y=L_y/2$ for the model with spurious lattice effects and for the corrected model, in which discrete lattice effects are canceled at order $\dt$.}
\label{fig:corrections_div_u}
\end{figure}

In this section we validate the improvement of our method by the cancelation of discrete lattice effects, which arise from the forcing term of the lattice Boltzmann equation. As already mentioned in Sec. \ref{sec:discrete_lattice_effects}, discrete lattice effects manifest themselves as spurious source terms of order $\dt$ in the Navier-Stokes equations [see Eq. (\ref{eq:NS_discrete_lattice_effects})], which --- depending on the strength of the spatial deformation --- have a considerable effect on the fluid. To illustrate this effect, we have performed two-dimensional simulations \textit{with} and \textit{without} the spurious source terms of order $\dt$ for a curved medium with $8$ metric perturbations of amplitude $a_0 = 0.1$ and range $r_0 = 22$. The fluid is driven through the medium by an external force corresponding to a pressure drop of $\nabla P = 10^{-6}$ in $x$ direction, and the relaxation time is set to $\tau = 1$. 
Figure \ref{fig:paper_corrections_streamlines} shows the velocity field of the fluid for a simulation with discrete lattice effects (\textit{left}) and the same simulation using the corrected method (\textit{right}). 
In the left figure, one can clearly see large unphysical sinks in the velocity field, originating from the spurious source terms in the Navier-Stokes equations. By properly correcting the density and velocity fields in each time step as described in Sec. \ref{sec:discrete_lattice_effects}, these unphysical source terms are eliminated at order $\dt$ , leading to the source-free velocity field depicted in the right figure, which appears much more physical.
The improvement becomes even more obvious in the corresponding divergence field of the velocity, which is plotted as a function of $x$ in Fig. \ref{fig:corrections_div_u} at the cross-section $y = L_y/2$. 
According to the continuity Eq. (\ref{eq:NS}), $\cov_i (\rho u^i)$ must vanish identically in the stationary state, which is clearly fulfilled for the corrected method, whereas the uncorrected method causes large deviations due to the unphysical source terms.

\subsection{Finite resolution study}\label{sec:finite-resolution-study}

\begin{figure}[ht]
\includegraphics[width=0.8\columnwidth]{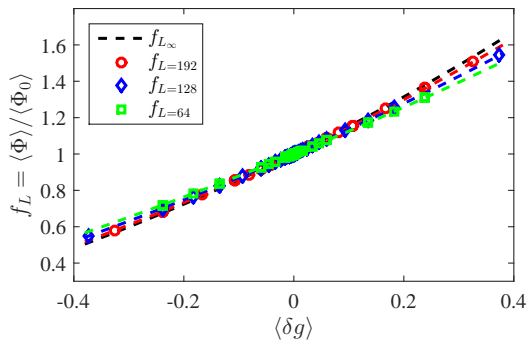}
\caption{Flux ratio $\langle \Phi \rangle / \langle \Phi_0 \rangle$ as function of  $\langle \dg \rangle$ for different system sizes.}
\label{fig:paper_flux_finite_size}
\end{figure}

\begin{figure}[ht]
\includegraphics[width=0.8\columnwidth]{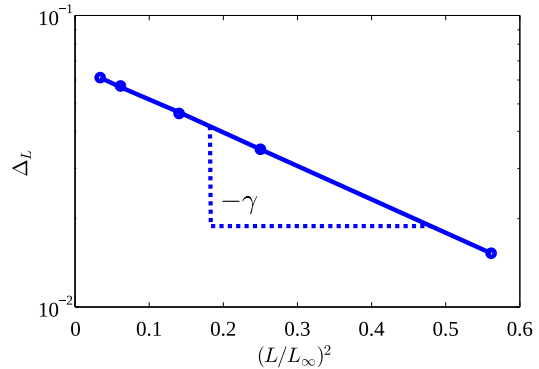}
\caption{Relative distance between the flux curves as function of the system size $L^2$.}
\label{fig:paper_flux_finite_size1}
\end{figure}

In order to investigate the effect of the grid resolution on our data we have compared simulations of the two-dimensional curved medium described in Sec. \ref{sec:campylotic-media-regular}) for different grid resolutions $\dx$, corresponding to different system lengths $L = 128/\dx$. 
Figure \ref{fig:paper_flux_finite_size} depicts the corresponding flux curves for system lengths $L = 64$, $128$, and $192$, where we abbreviate the flux ratio at system length $L$ by $f_L := \langle \Phi \rangle / \langle \Phi_0 \rangle$. 
As can be seen, for increasing system lengths $L$ the flux curves $f_L$ converge rapidly to the limiting function $f_{L_\infty}$, which we approximate by the flux law given by Eq. (\ref{eq:flux-law-2D}) for $L=256$. In order to prove the convergence quantitatively, we show that the relative distance between the curves, defined as
\begin{align*}
	\Delta_L^2 := \frac{\left\| f_{L} - f_{L_\infty} \right\|^2}{\left\| f_{L_\infty} \right\|^2}
	= \frac{\D\int \left| f_{L}(x) - f_{L_\infty}(x) \right|^2 dx}{\D\int \left|  f_{L_\infty}(x) \right|^2 dx},
\end{align*} 
converges to zero with increasing system size $L^2$. Figure \ref{fig:paper_flux_finite_size1} depicts the relative distance $\Delta_L$ as function of the system size in a semi-logarithmic plot. 
As can be seen, the flux curves $f_L$ converge to $f_{L_\infty}$ exponentially with $L^2$,
\begin{align*}
	\Delta_L \sim \exp \left( -\gamma\, \frac{L^2}{L_\infty^2} \right),
\end{align*}
with $\gamma = 2.66 \pm 0.05$. Consequently, we conclude that --- within an error of about $1 \%$ --- our numerical results are free of finite resolution effects.

\subsection{Flux conservation}\label{sec:flux-conservation}

\begin{figure}[ht]
\includegraphics[width=0.8\columnwidth]{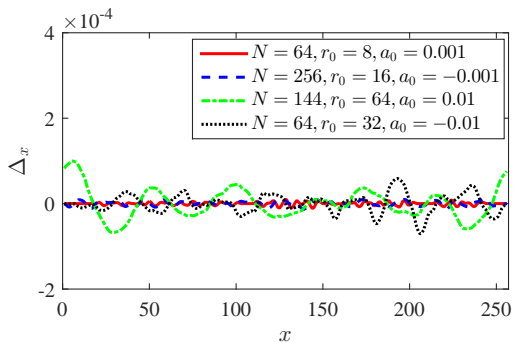}
\caption{Relative fluctuation of the flux, $\Delta_x = (\Phi - \langle \Phi \rangle_x)/\langle \Phi \rangle_x$, as function of $x$.}
\label{fig:paper_flux_vs_x}
\end{figure}

In order to check to which extent the flux is numerically conserved, it is insightful to plot the profiles $\Phi(x)$ along the channel direction $x$. In Fig. \ref{fig:paper_flux_vs_x}, the relative fluctuation of the flux around its mean value, $\Delta_x := (\Phi - \langle \Phi \rangle_x)/\langle \Phi \rangle_x$, is plotted versus $x$ for different randomly curved media. As can be seen, the flux varies only marginally with $x$, with an average standard deviation of $\Delta_x \approx 2 \cdot 10^{-5}$. Thus, we conclude that the flux is numerically conserved up to a negligible error of about $0.002 \%$. 

The conservation of the flux is closely connected to the improvements of the method depicted in Fig. \ref{fig:corrections_div_u}, since the flux loss between two channel cross-sections $x=x_1$ and $x=x_2$ is proportional to the divergence of $(\rho u^i)$:
\begin{align*}
	\Phi(x_2) - \Phi(x_1)
	= \oint_{\del V} g\left(\rho \vec u, \vec n \right) d(\del V)
	= \int_V \nabla_i (\rho u)^i dV,
 \end{align*}
where $V$ denotes the volume between the channel cross-sections. Here, we have used Gauss' integration theorem and the fact that the flux vanishes at the walls.

\section{Conclusions}

In this paper, we have studied Poiseuille flow in curved space, using the lattice Boltzmann method in curved space to simulate the flow according to the covariant Navier-Stokes equations. We have established a relation between the flux of the Poiseuille flow in the stationary state and the metric parameters of the medium (e.g., amplitude $a_0$, range $r_0$, and density $n$ of the metric perturbations) for different types of curved media, starting from two-dimensional media with regularly and randomly arranged metric perturbations and ending with three-dimensional media. In all cases, we have found that the flux depends only on a specific combination of parameters, which we have identified as the average metric perturbation,
\begin{align*}
	\langle \dg \rangle \sim a_0\, r_0^D\, n,
\end{align*}
where $D$ denotes the dimension of the medium.
Even beyond leading order in the metric perturbations, the flux is well-described as a function of the average metric perturbation $\langle \dg \rangle$. Explicitly, we have found that the (spatially averaged) flux $\langle \Phi \rangle$ is well-described by the following flux law:
\begin{align*}
	\langle \Phi \rangle = \langle \Phi_0 \rangle \cdot 
	\left( 1 + \a^D \cdot \langle \dg \rangle 
	 + D \a^D \b \cdot \langle \dg \rangle^2 \right),
\end{align*}
where $\langle \Phi_0 \rangle$ denotes the (spatially averaged) flux in flat space and $\a$ and $\b$ are fitting parameters. This relation is universal in the sense that --- within the parameter range considered here --- it depends  neither on the specific shape nor on the spatial order of the metric perturbations and is valid both for two- and three-dimensional systems. 


For the simulations, we have improved our curved-space lattice Boltzmann method further by canceling discrete lattice effects, which originate from the forcing term, at order $\dt$. These discrete lattice effects appear as spurious source terms in the Navier-Stokes equations and thus reduce the accuracy of the method. In curved spaces, the cancelation of spurious terms turns out to be a sophisticated task since the inertial forces depend quadratically on the fluid velocity and even contribute to the continuity equation. Still, it has been possible to eliminate all spurious terms of order $\dt$ by correcting not only the fluid velocity at each time step (as is commonly suggested in the literature \cite{Guo2002}), but also the fluid density. With the corrected model at hand, we have been able to improve the accuracy of the results considerably.

\appendix
\section{Cancelation of Discrete Lattice Effects: A Chapman-Enskog Analysis}\label{sec:app}

In the following derivation, we will use the abbreviations
\begin{align*}	
	\sumvar_\l := {\sum}_\l \sqrt g 	
	\quad , \quad
	D_t := \del_t + c_\l^i \del_i
	\quad , \quad
	\delvar_i := \del_i - \G^j_{ij}.
\end{align*}

\subsection{Ansatz and preparation}\label{sec:app1}

We show that spurious discrete lattice effects originating from the forcing term can be canceled at first order in $\dt$ by a redefinition of density $\r \rightarrow \rr$ and velocity $u^i \rightarrow \uu^i$. Inspired by Refs. \cite{Guo2002,li2010}, we make the following ansatz for the corrected quantities $\rr$ and $\uu^i$:
\begin{align}
	\label{eq:CE2-redefinitons1}
	\rr &= \leer{$u^i$}\r + \dt \,R(\rr,\uu) = \sumvar_\l  f_\l  + \dt \,R(\rr,\uu), \\
	\label{eq:CE2-redefinitons2}
	\rr \uu^i &= \r u^i + \dt \,U^i(\rr,\uu) = \sumvar_\l f_\l c^i_\l  + \dt\, U^i(\rr,\uu),
\end{align}
where $R(\rr,\uu)$ and $U^i(\rr,\uu)$ are correction functions whose explicit form has to be determined in such a way that discrete lattice effects are canceled in the Navier-Stokes equations. The corrected quantities $\rr$ and $\uu^i$ are determined implicity by Eqs. (\ref{eq:CE2-redefinitons1}) and (\ref{eq:CE2-redefinitons2}). This system of equations   can be solved for $\rr$ and $\uu^i$ by using standard numerical techniques (e.g., Newtons algorithm, see Appendix \ref{sec:newton_algorithm}). To find the explicit expressions for the correction functions $R(\rr,\uu)$ and $U^i(\rr,\uu)$, we perform a Chapman-Enskog expansion of the lattice Boltzmann equation in curved space,
\begin{align}\label{eq:CE2-LB}
	f_\l(x^i + c_\l^i \dt, t + \dt) - f_\l(x^i, t) 
	= - \frac{1}{\t} \left( f_\l - f_\l^{\rm eq} \right)
	+ \dt \mathcal F_\l.
\end{align}  
The equilibrium distribution as well as the forcing term are evaluated in terms of the corrected quantities $\rr$ and $\uu$, i.e., $f^{\rm eq}_\l = f^{\rm eq}_\l(\rr,\uu)$, $\FF_\l = \FF_\l(\rr,\uu)$. Both are expanded into tensor Hermite polynomials,
\begin{align*}
	\HH_{(0)} &= 1, \quad
	\HH_{(1)}^i = \frac{c_\l^i}{c_s}, \quad
	\HH_{(2)}^{ij} = \frac{c_\l^i c_\l^j}{c_s^2} - \d^{ij}, \\
	\HH_{(3)}^{ijk} &= \frac{c_\l^i c_\l^j 
	c_\l^k}{c_s^3} - \left( \delta^{ij} \frac{c_\l^k}{c_s} + \delta^{jk}
	\frac{c_\l^i}{c_s} + \delta^{ki} \frac{c_\l^j}{c_s} \right).
\end{align*}
This yields (up to second order):
\begin{align}
	\label{eq:CE2-feq-expansion}	
	f_\l^{\rm eq} &= \frac{w_\l}{\sqrt g} 
	\sum_{n=0}^3 \frac{1}{n! \, c_s^n} a_{(n)}^{{\rm eq},I_n} \, \HH_{(n)}^{I_n},
\end{align}
where $I_n = (i_1,\ldots,i_n)$ denotes an index $n$-tuple, which is summed over. The expansion coefficients are given by
\begin{align*}
	&a_{(0)}^{{\rm eq}} = \rr,\qquad
	a_{(1)}^{{\rm eq},i} = \rr \uu^i,\qquad 
	a_{(2)}^{{\rm eq},ij} = \rr c_s^2 \Delta^{ij} + \rr \uu^i \uu^j, \\
	&a_{(3)}^{{\rm eq},ijk} = \rr c_s^2 \left( \Delta^{ij} \uu^k + \Delta^{jk} \uu^i +
	\Delta^{ki} \uu^j \right) + \rr \uu^i \uu^j \uu^k,
\end{align*}
where $\Delta^{ij} := g^{ij} - \delta^{ij}$ is a measure for the deviation from flat space.
The forcing term reads
\begin{align}
	\label{eq:CE2-F-expansion}
	\mathcal F_\l &= \frac{w_\l}{\sqrt g} 
	\sum_{n=0}^2 \frac{1}{n! \, c_s^{n+1}}\, a_{(n)}^{I_n} F_\l^{i_{n+1}} \, \HH_{(n+1)}^{I_{n+1}},
\end{align}
with coefficients
\begin{align*}
	a_{(0)} = a_{(0)}^{{\rm eq}}, \qquad
	a_{(1)}^{i} = a_{(1)}^{{\rm eq},i},\qquad
	a_{(2)}^{ij} = a_{(2)}^{{\rm eq},ij} - \s^{ij}.
\end{align*}
Here, $F_\l^i = - \Gamma^i_{jk} c_\l^j c_\l^k$ represents inertial forces and $\s^{ij} = -( 1 - \frac{1}{2\t}) \sumvar_\l c_\l^i c_\l^j \left( f_\l - f^{{\rm eq}}_\l \right) $ is the viscous stress tensor.

From Eqs. (\ref{eq:CE2-feq-expansion}) and (\ref{eq:CE2-F-expansion}), we can calculate the moments of $f^{\rm eq}_\l$ and $\FF_\l$, which are needed for the Chapman-Enskog expansion later. For this purpose, we apply the orthogonality condition of the Hermite polynomials, which is supported by the D3Q41 lattice up to third order,
\begin{align*}
	\sum_\l w_\l \HH_{(n)}^{I_n} \HH_{(m)}^{J_m} &= \d_{nm} \d^{I_n J_n},
	\qquad 0 \leq n,m \leq 3,
\end{align*}
where $I_n = (i_1,\ldots,i_n), J_m = (j_1,\ldots,j_m)$ are index tupels and $\d^{I_n J_n} := \sum_{\s \in S_n}(\d^{i_1 j_{\s(1)}} \cdots \d^{i_n j_{\s(n)}})$.

This yields:
\begin{align}
	\label{eq:CE2-eq-moments1}
	\rr &= \sumvar_\l f^{\rm eq}_\l, \\	
	\label{eq:CE2-eq-moments2}
	\rr \uu^i &= \sumvar_\l f^{\rm eq}_\l c_\l^i, \\
	\label{eq:CE2-eq-moments3}
	\Pi^{{\rm eq},ij} &= \sumvar_\l f^{\rm eq}_\l c_\l^i c_\l^j 
	= \rr\left( c_s^2 g^{ij} + \uu^i \uu^j \right),\\
	\nonumber
	\Sigma^{{\rm eq},ijk} &= \sumvar_\l f^{\rm eq}_\l c_\l^i c_\l^j c_\l^k 
	= \rr c_s^2 \left( \uu^i g^{jk} + \uu^j g^{ik} + \uu^k g^{ij} \right)\\
	\label{eq:CE2-eq-moments4}
	&\qquad\qquad\qquad\qquad + \OO(\uu^3) \\
	\label{eq:CE2-A}
	A &= \sumvar_\l \FF_\l 
	= - \left( \G^i_{ij} \rr \uu^j + \G^i_{ji} \rr \uu^j \right), \\
	\label{eq:CE2-B}
	B^i &= \sumvar_\l \FF_\l c_\l^i 
	= - \left( \G^i_{jk} T^{jk} + \G^j_{jk} T^{ki}  + \G^j_{kj} T^{ki} \right),\\
	\nonumber
	C^{ij} &= \sumvar_\l \FF_\l c_\l^i c_\l^j 
	= -\G^i_{kl} \Sigma^{{\rm eq},jkl}
	-\G^j_{kl} \Sigma^{{\rm eq},ikl}\\
	\label{eq:CE2-C}
	&\qquad\qquad\qquad\quad
	-\G^k_{kl} \Sigma^{{\rm eq},ijl}
	-\G^l_{kl} \Sigma^{{\rm eq},ijk},
\end{align}
Here, we have defined $T^{ij} := \Pi^{{\rm eq},ij} - \s^{ij}$ and neglected terms of order $\OO(\uu^3) \sim \OO(\text{Ma}^3)$ in $C^{ij}$, where $\text{Ma}$ denotes the Mach number.

\subsection{Multiscale Expansion}

We start with the Chapman-Enskog expansion by expanding the distribution function and the time derivative in the Knudsen number $\e$:
\begin{align*}
	f &= f^{(0)} + \e f^{(1)} + \e^2 f^{(2)} + ... , \\
	\del_t &= \e \del_t^{(1)} + \e^2 \del_t^{(2)} + ... , \\
\end{align*}
Furthermore, we rescale all other quantities $Q = (\del_i, \cov_i, \G^i_{jk}, \FF, A, B^i, R, U^i, \s^{ij}, ...)$ by the Knudsen number, $Q = \e\, Q^{(1)}$.

Plugging everything into Eq. (\ref{eq:CE2-LB}) and comparing orders of $\e$, we obtain the following equations:
\begin{align}
	\label{eq:CE2-0}
	\T\OO(\e^0): \qquad&
	f_\l^{(0)} = f_\l^{{\rm eq}}, \\ 
	\label{eq:CE2-I}
	\T\OO(\e^1): \qquad&
	\T D_t^{(1)} f_\l^{(0)} = - \frac{1}{\t \dt} f_\l^{(1)} + \FF_\l^{(1)},\\
	\nonumber
	\T\OO(\e^2): \qquad&
	\T\del_t^{(2)} f_\l^{(0)} + \left(1- \frac{1}{2\t}\right) D_t^{(1)} f_\l^{(1)} \\
	\label{eq:CE2-II}
	&\T= - \frac{1}{\t \dt} f_\l^{(2)} - \frac{\dt}{2} D_t^{(1)} \FF_\l^{(1)}.
\end{align}

Comparing Eqs. (\ref{eq:CE2-redefinitons1}) and (\ref{eq:CE2-redefinitons2}) to Eqs. (\ref{eq:CE2-eq-moments1}) and (\ref{eq:CE2-eq-moments2}) with $f_\l^{(0)} = f_\l^{{\rm eq}}$, we see that
\begin{align}	
	\label{eq:CE2-moments3}
	\sumvar_\l f^{(1)}_\l  = - \dt\, R^{(1)},
	\qquad 
	\sumvar_\l f^{(1)}_\l c_\l^i  = - \dt\, U^{(1),i}.
\end{align}

\subsection{Moments of Eqs. (\ref{eq:CE2-I}) and (\ref{eq:CE2-II})}

Taking the moments of Eq. (\ref{eq:CE2-I}) yields:
\begin{align}
	\label{eq:CE2-1}
	\T\sumvar_\l [\text{Eq. (\ref{eq:CE2-I})}]  :
	\T\quad &\del_t^{(1)} \rr + \delvar_i^{(1)} \left(\rr \uu^i \right)
	\T= A^{(1)} + \frac{1}{\t} R^{(1)}, \\
	\label{eq:CE2-2}
	\T\sumvar_\l c_\l^i\, [\text{Eq. (\ref{eq:CE2-I})}]  : \quad &
	\T\del_t^{(1)} \left( \rr \uu^i \right) + \delvar_j^{(1)} \Pi^{(0),ij}
	\T= B^{(1),i} + \frac{1}{\t} U^{(1),i},
\end{align}
where $\Pi^{(0),ij} = \Pi^{{\rm eq},ij} = \rr\left( c_s^2 g^{ij} + \uu^i \uu^j \right)$, and the $\delvar_i$ derivative originates from
\begin{align*}
	{\sum}_\l \sqrt g\, c_\l^i \del_i f_\l^{(0)}
	&= \del_i {\sum}_\l \sqrt g\, c_\l^i f_\l^{(0)}
	-{\sum}_\l  (\del_i \sqrt g)\, c_\l^i f_\l^{(0)}\\
	&= \del_i (\rr \uu^i) - \G^j_{ij} (\rr \uu^i) =: \delvar_i (\rr \uu^i),
\end{align*}
where we have used the identity $\del_i \sqrt g = \G^j_{ij} \sqrt g$.
The moments of Eq. (\ref{eq:CE2-II}) are given by
\begin{align}
	\nonumber
	&\T\sumvar_\l [\text{Eq. (\ref{eq:CE2-II})}] :\\
	\nonumber
	&\T\del_t^{(2)} \rr 
	\T= \dt \left( 1 - \frac{1}{2\t}\right) \left( \del_t^{(1)} R^{(1)} +  \delvar_i^{(1)} U^{(1),i}\right) \\
	\label{eq:CE2-3}	
	&\qquad\qquad\quad\quad\ \T-\frac{\dt}{2} \left( \del_t^{(1)} A^{(1)} +  \delvar_i^{(1)} B^{(1),i} \right), \\	
	\nonumber
	&\T\sumvar_\l c_\l^i\, [\text{Eq. (\ref{eq:CE2-II})}] :\\	
	\nonumber
	&\T\del_t^{(2)} \left(\rr \uu^i \right) 
	\T= \delvar_j^{(1)} \s^{(1),ij}-\frac{\dt}{2} \del_t^{(1)} B^{(1),i} \\
	\label{eq:CE2-4}
	&\qquad\qquad\qquad\T+ \dt \left( 1 - \frac{1}{2\t}\right) \del_t^{(1)} U^{(1),i},
\end{align}
where $\s^{(1),ij}$, the viscous stress tensor (rescaled by $\e$), is defined as
\begin{align}\label{eq:CE-stress}
	\T\s^{(1),ij} = - \left(1 - \frac{1}{2\t}\right) \sumvar_\l f^{(1)}_\l c_\l^i c_\l^j .
\end{align}
The explicit expression for the viscous stress tensor will be derived later.

\subsection{Continuity Equation}

For the continuity equation, we add $\e [\text{Eq. (\ref{eq:CE2-1})}]$ and $\e^2 [\text{Eq. (\ref{eq:CE2-3})}]$:
\begin{align}
	\nonumber	
	& \T\quad\del_t \rr +  \delvar_i \left(\rr \uu^i \right)
	= A + \frac{1}{\t} R 	
	-\frac{\dt}{2} \left( \del_t A + \delvar_i B^i \right) \\
	\label{eq:CE2-5}
	& \T\qquad\qquad\qquad\quad\quad+ \dt \left( 1 - \frac{1}{2\t}\right) \left( \del_t R +  \delvar_i U^i \right). 
\end{align}
In order to cancel all spurious terms of order $\dt$, we first rescale the forcing term by a factor of $\left( 1 - \frac{1}{2\t}\right)$,
\begin{align}
	\label{eq:CE2-F1}
	\T\T\FF_\l \rightarrow \left( 1 - \frac{1}{2\t}\right) \FF_\l,
\end{align}
which implies $(A,B^i) \rightarrow \left( 1 - \frac{1}{2\t}\right) (A,B^i)$. With this redefinition, Eq. (\ref{eq:CE2-5}) becomes
\begin{align}
	\nonumber
	& \quad\del_t \rr + \delvar_i \left(\rr \uu^i \right)
	= \T\left( 1 - \frac{1}{2\t}\right) A + \frac{1}{\t} R \\
	\label{eq:CE2-5a}
	&+ \T\dt \left( 1 - \frac{1}{2\t}\right) \left( \del_t \left( R -\frac{1}{2} A \right)  + \delvar_i \left( U^i - \frac{1}{2} B^i \right) \right). 
\end{align}
Finally, to cancel all spurious terms of order $\dt$, we set 
\begin{align}
	\label{eq:CE2-ST}
	R = \frac{1}{2} A, \qquad\qquad
	U^i = \frac{1}{2} B^{{\rm eq},i},
\end{align}
where $B^{{\rm eq},i}$ denotes the equilibrium part of $B^{i}$ given by
\begin{align*}
B^{{\rm eq},i} = - \G^i_{jk} \Pi^{{\rm eq},jk} - \G^j_{jk} \Pi^{{\rm eq},ki} - \G^j_{kj} \Pi^{{\rm eq},ki}.
\end{align*}
[Note that for second-order accuracy in $\dt$, it is sufficient to cancel the equilibrium part of $\dt B^i$ in Eq. (\ref{eq:CE2-5a}), since the non-equilibrium part, $\dt B^{{\rm neq}, i}$, is of order $\OO(\dt^2)$.]
With this choice of $R$ and $U^i$, Eq. (\ref{eq:CE2-5a}) becomes
\begin{align*}	
	\quad\del_t \rr + \delvar_i \left(\rr \uu^i \right) = A + \OO(\dt^2),
\end{align*}
which, after inserting the explicit expression for $A$ [Eq. (\ref{eq:CE2-A})], yields the correct continuity equation at order $\OO(\dt^2)$:
\begin{align*}
	\del_t \rr + \cov_i \left(\rr \uu^i \right)	&= 0,
\end{align*}
where $\cov$ denotes the covariant derivative.

\subsection{Momentum Equation}

Adding $\e [\text{Eq. (\ref{eq:CE2-2})}]$ and $\e^2 [\text{Eq. (\ref{eq:CE2-4})}]$ yields the momentum conservation equation:
\begin{align*}
	&\T \qquad\del_t \left( \rr \uu^i \right) 
	+ \delvar_j \Pi^{(0),ij}
	\T= \delvar_j \s^{ij} + B^i + \frac{1}{\t} U^i  \\
	&\T\qquad\qquad\qquad\qquad\qquad\quad-\frac{\dt}{2} \del_t B^i	+ \dt \left( 1 - \frac{1}{2\t}\right) \del_t U^i.
\end{align*}
Applying the same rescaling of the forcing term as before, $(A,B^i) \rightarrow \left( 1 - \frac{1}{2\t}\right) (A,B^i)$, together with the constraints $(R,U^i) = \frac{1}{2} (A,B^i)$ , the momentum equation simplifies to
\begin{align*}
	\del_t \left( \rr \uu^i \right) + \delvar_j \Pi^{(0),ij}
	= \delvar_j \s^{ij} + B^i.
\end{align*}
Inserting the explicit expression for $B^i$ [Eq. (\ref{eq:CE2-B})] and $\Pi^{(0),ij} = \Pi^{{\rm eq},ij}$ [Eq. (\ref{eq:CE2-eq-moments3})] yields the familiar Navier-Stokes equation:
 \begin{align*}
	\del_t \left( \rr\, \uu^i \right) + \cov_j \left( \rr\, \uu^i \uu^j + \rr \,c_s^2 g^{ij} \right) 
	&= \cov_j \s^{ij},
\end{align*}
where $\s^{ij}$ denotes the viscous stress tensor, whose explicit form in terms of $\rr$ and $\uu^i$ can be derived as follows:
\begin{align*}
	\s^{ij} \overset{(\ref{eq:CE-stress})}=&\T - \left(1 - \frac{1}{2\t}\right) \e\, \sumvar_\l c_\l^i c_\l^j f_\l^{(1)}\\
	\overset{(\ref{eq:CE2-I})}=& \T\left( \t - \frac{1}{2}\right) \dt \,\e \sumvar_\l \left( c_\l^i c_\l^j D_t^{(1)} f_\l^{(0)} - \FF_\l^{(1)} \right) \\
   \approx\ \ & \T \left( \t - \frac{1}{2}\right) \dt \left( \delvar_k \Sigma^{{\rm eq},ijk} - C^{ij} \right),
\end{align*}
where we have assumed that $\left( \t - \frac{1}{2} \right) \dt\, \del_t^{(1)} \Pi^{(0),ij} \ll 1$.
After inserting the explicit expressions for $\Sigma^{{\rm eq},ijk}$ [Eq. (\ref{eq:CE2-eq-moments4})] and $C^{ij}$ [Eq. (\ref{eq:CE2-C})], we obtain
\begin{align*}
	\s^{ij} &=  \T \nu \left( \cov^j (\rr \uu^i) + \cov^i (\rr \uu^j)  + g^{ij} \cov_k (\rr \uu^k)  \right),
\end{align*}
where we have defined $\nu := \left( \t - \frac{1}{2}\right) \dt\, c_s^2$ and neglected terms of the order $\OO(\uu^3) \sim \OO(\text{Ma}^3)$, $\text{Ma}$ being the Mach number.

\subsection{Newton's algorithm}\label{sec:newton_algorithm}

The corrected density $\rr$ and velocity $\uu^i$ are determined implicitly by the coupled Eqs. (\ref{eq:CE2-redefinitons1}) and (\ref{eq:CE2-redefinitons2}):
\begin{align*}
	\rr &= \r  + \dt\, R(\rr,\uu), \\
	 \rr \uu^i &= \r u^i + \dt\, U^i(\rr,\uu),
\end{align*}
where $R$ and $U^i$ are given by Eqs. (\ref{eq:CE-summary1}) and (\ref{eq:CE-summary2}). Since  $R(\rr,\uu)$ and $U(\rr,\uu)$ depend only linearly on $\rr$, we can eliminate the density by defining $\RR(\uu) := R(\rr,\uu)/\rr$ and $\UU(\uu) := U(\rr,\uu)/\rr$. Now, we can rewrite Eqs. (\ref{eq:CE2-redefinitons1}) and (\ref{eq:CE2-redefinitons2}) as follows:
\begin{align}
	\label{eq:CE2-newton}
	\frac{\r}{\rr} &= 1 - \dt\, \RR(\uu), \\
	\nonumber
	\uu^i &= \frac{\r}{\rr} u^i + \dt\, \UU^i(\uu).
\end{align}
Inserting the first equation into the second, we obtain an equation for $\uu$ which is decoupled from the density $\rr$:
\begin{align*}
	\uu^i &= \left( 1 - \dt\, \RR(\uu) \right) u^i + \dt\, \UU^i(\uu).
\end{align*}
After defining
\begin{align*}
	G^i(\uu) := \left( 1 - \dt\, \RR(\uu) \right) u^i + \dt\, \UU^i(\uu) - \uu^i,
\end{align*}
Newton's algorithm can be used to solve $G^i(\uu) = 0$. To this end, one assigns initial conditions $\uu^i_{(0)} = u^i$ and iterates,
\begin{align*}
	\uu^i_{(n+1)} = \uu^i_{(n)} - \left[ J^{-1} \right]^i_j G^j(\uu_{(n)}),
\end{align*}
where $J = (\frac{dG^i}{d\uu^j})$ denotes the Jacobian matrix.
This procedure converges rapidly, and typically after 1--3 iterations, a sufficiently accurate solution is found. Given the solution for $\uu$, the solution for $\rr$ is obtained from Eq. (\ref{eq:CE2-newton}).

\begin{acknowledgements}
  We acknowledge financial support from the European Research Council (ERC) Advanced
  Grant No. 319968-FlowCCS.
\end{acknowledgements}

\bibliography{citations.bib}
\end{document}

%% file: pakete.tex
\usepackage[T1]{fontenc}
\usepackage[latin1]{inputenc}

\usepackage[british]{babel}
	\addto\captionsngerman{}

\usepackage{amsmath,amssymb,amsthm,dsfont,mathtools,mathrsfs,bbm}

\usepackage{mdwlist}
\usepackage{paralist}
\usepackage{array}
\usepackage{booktabs}						
\usepackage{dcolumn} 						

\usepackage{graphicx}
\usepackage{subfigure}  
\usepackage{wrapfig}

\usepackage{url}

\usepackage{color}
	\definecolor{myblue}{rgb}{0,0.3,0.8}
	\definecolor{mygreen}{rgb}{0,0.5,0}
	\definecolor{myblue}{rgb}{0,0.3,0.8}

\usepackage{hyperref} 
	\hypersetup{
		colorlinks, 
		linkcolor = {myblue}, 
		citecolor = {mygreen}, 
		urlcolor  = {myblue}
	}

\usepackage{savesym}

\usepackage{pdfpages}


%% file: befehle.tex
\savesymbol{a}
\savesymbol{b}
\savesymbol{d}
\savesymbol{e}
\savesymbol{k}
\savesymbol{l}
\savesymbol{H}
\savesymbol{L}
\savesymbol{p}
\savesymbol{r}
\savesymbol{s}
\savesymbol{S}
\savesymbol{F}
\savesymbol{t}
\savesymbol{w}
\savesymbol{th}
\savesymbol{div}
\savesymbol{tr}
\savesymbol{Re}
\savesymbol{Im}
\savesymbol{AA}
\savesymbol{SS}
\savesymbol{Si}

\newcommand{\a}{\alpha}
\newcommand{\b}{\beta}
\newcommand{\e}{\varepsilon}
\newcommand{\d}{\delta}

\newcommand{\G}{\Gamma}
\newcommand{\l}{\lambda}

\newcommand{\r}{\rho}
\newcommand{\s}{\sigma}
\newcommand{\t}{\tau}

\newcommand{\dt}{{\Delta t}}
\newcommand{\dx}{{\Delta x}}
\newcommand{\dy}{{\Delta y}}
\newcommand{\dz}{{\Delta z}}
\newcommand{\dg}{{\delta g}}

\newcommand{\FF}{{\mathcal F}}
\newcommand{\OO}{{\mathcal O}}

\newcommand{\HH}{{\mathcal H}}
\newcommand{\RR}{{\mathcal R}}
\newcommand{\UU}{{\mathcal U}}
\newcommand{\cov}{{\nabla}}
\newcommand{\rr}{{\hat\rho}}
\newcommand{\uu}{{\hat u}}

\newcommand{\del}{\partial}
\newcommand{\delvar}{{\overline \partial}}
\newcommand{\sumvar}{{\overline \sum}}

\newcommand{\Z}{\mathbbm{Z}}

\newcommand{\pdiff}[2]{\frac{\partial#1}{\partial #2}}

\newcommand{\D}{\displaystyle}
\newcommand{\T}{\textstyle}

%% file: makros.tex
\newlength{\tmplen}
\newcommand{\leer}[1]{\settowidth{\tmplen}{#1}\hspace*{\tmplen}}

